\def\lesssim{{_ <\atop{^\sim}}}
\def\grtsim{{_ >\atop{^\sim}}}
\def\apj{ApJ}
\def\aj{AJ}

\def\mnras{MNRAS}
\def\aap{A\&A}
\def\apjS{ApJS}
\documentstyle[twocolumn,epsfig]{mn}

\title[Disc galaxy evolution models in a hierarchical formation scenario]
{Disc galaxy evolution models in a hierarchical formation 
scenario: structure and dynamics}

\author[Firmani \& Avila-Reese]
       {Claudio Firmani,$^{1,2,3,4}$
        \newauthor
        Vladimir Avila-Reese, $^{1,3,4}$\\
$^1$Instituto de Astronom\'\i a, UNAM, A.P. 70-264, 04510 M\'exico D.F., M\'exico\\ 
$^2$Osservatorio Astronomico di Brera, via E.Bianchi 46, I-23807 Merate, Italy \\
$^3$Also  Astronomy Department, New Mexico State University, Box 30001, Dept.
4500, Las Cruces, NM 88003-0001\\
$^4$Email: firmani, avila@astroscu.unam.mx}

   \date{Received ...; accepted ...}

\begin{document}

\maketitle

\begin{abstract}
We predict the internal structure and dynamics of 
present-day disc galaxies using galaxy evolution models within a 
hierarchical formation scenario. The halo mass aggregation histories, 
for a flat cold dark matter model with
cosmological constant, were generated and used to calculate the 
virialization of dark matter haloes. A diversity of halo density 
profiles were obtained, the most typical one being close to 
that suggested by Navarro, Frenk \& White. We modeled 
the way in which discs in centrifugal equilibrium are build 
within the evolving dark haloes, using gas accretion 
rates proportional to the halo mass aggregation rates, 
and assuming detailed angular momentum conservation. We calculated:
the gravitational interactions between halo and disc ---including 
the adiabatic contraction of the halo due to disc formation---, 
and the hydrodynamics, star formation, and evolution of the galaxy 
discs. We find that the slope and zero-point of the 
Tully-Fisher (TF) relation in the infrared bands may be explained as a
direct consequence of the cosmological initial conditions. This relation is
almost independent of the assumed disc mass fraction, when the disc 
component in the rotation curve decomposition is non-negligible. 
Thus, the power spectrum of fluctuations can be normalized at 
galaxy scales through the TF relation independently of the disc 
mass fraction assumed.
The rms scatter of the model TF relation originates mainly from the 
scatter in the dark halo structure and, to a minor extension, from 
the dispersion of the primordial spin parameter $\lambda$. The scatter 
obtained from our models does not disagree with the 
observational estimates. Our models
allow us to understand why the residuals of the TF relation do not
correlate significantly with disc size or surface brightness. We can
also explain why low and high surface brightness galaxies have 
the same TF relation; the key point is the dependence of 
the star formation efficiency on the disc surface density.
The correlations between gas fraction and surface brightness, and 
between scale length and $V_{\max}$ obtained with our models agree
with those observed. Discs formed within the growing haloes, where
$\lambda$ is assumed to be time independent, have nearly exponential surface
density distributions. The shape of the rotation curves changes with
disc surface brightness and is nearly flat for most cases. The rotation
curve decompositions show a dominance of dark matter down to very 
small radii, in conflict with some observational inferences.
The introduction of shallow cores in the dark halo attenuates this difficulty
and produces haloes with slightly smaller rotation velocities. Other features of
our galaxy models are not strongly influenced by the shallow core.

\end{abstract}


\section{Introduction}

The understanding of galaxy formation and evolution provides
a crucial link between a large
body of astronomical observations and cosmological theories. 
According to the inflationary cold dark matter (CDM) theory, 
large-scale cosmic structures arise from primordial density fluctuations that
evolve gravitationally through a hierarchical process of mass
aggregation (accretion and merging). In order to explain
galaxy formation, it is also necessary to take into account the
hydrodynamical and dissipative processes related with the baryon matter,
as well as the poorly understood processes of star 
formation (SF) and its feedback. In 
this paper we use galaxy evolution models, 
which are able to predict the local and global properties of disc galaxies,
and integrate them into the cosmological framework.

In the last decade, several galaxy formation and evolution models
based on semi-analytical and analytical approaches were developed (e.g. White
\& Frenk 1991; Lacey \& Silk 1993; Kauffmann, White \& Guiderdoni 1993; 
Cole et al. 1994; Baugh, Cole \& Frenk 1996;  Somerville \& 
Primack 1998; Fall \& Efstathiou 1980; Dalcanton, Spergel \& 
Summer 1997; Mo, Mao \& White 1998, hereafter MMW98; van den Bosch 1998).
The models we present in this paper make use of some of the 
ingredients of these approaches. However, as opposed to most of the 
previous models, we calculate the {\it internal} structure, 
dynamics, hydrodynamics and SF process that define the evolution of 
a given galaxy (dark+luminous). That is why our approach is based on 
models of disc galaxy evolution
(Firmani, Hern\'{a}ndez \& Gallagher 1996). We link the initial 
and boundary conditions of these models to the cosmological setting through the
hierarchical mass aggregation histories (MAHs) and the structure of 
the dark matter (DM) haloes. We relax the simplifying 
hypothesis of a universal DM halo density profile and model 
the evolution of disc galaxies in a wide collection of DM structures. 
Thus, we can explore whether the initial
conditions given by the CDM models are consistent with the disc
galaxy properties and their correlations in the local universe.
We have carried out a sequence of {\it semi-numerical}
calculations, which include (i) the generation of the hierarchical MAHs from 
the primordial density fluctuation field, (ii) the gravitational
collapse and virialization of the DM haloes, (iii) the
formation of discs in centrifugal equilibrium within the evolving
haloes, (iv) the gravitational drag of the collapsing gas upon the DM 
halo, and (v) the evolution of a galaxy disc, including SF and
hydrodynamics. Some details were already disscused in
previous papers (Firmani et al. 1996; Avila-Reese, Firmani \& 
Hern\'{a}ndez 1998, hereafter AFH98; see also Avila-Reese 1998).

Our scenario suggests that discs form 
inside-out within a growing DM halo with a rate of gas accretion
proportional to the rate of cosmological mass aggregation 
(Gunn 1982; 1987; Ryden \& Gunn 1987; Avila-Reese \& Firmani 
1997; AFH98). We 
assume that the primordial angular momentum is acquired by the 
protogalaxy through tidal torques during its linear gravitational 
regime. This scenario will be called the ``extended collapse scenario'' 
as opposed to the ``merging scenario'' where the main properties of 
galaxies are a consequence of disc merging. Because the 
discs are dynamically frail objects (e.g. T\'oth \& Ostriker 1992), 
mergers could not have been relevant in establishing the main 
properties of disc galaxies, which constitute nearly the 80 per cent of normal
present-day galaxies. Compelling new observational
evidence suggests that a large fraction of baryon matter 
is in the form of small cold 
gas clumps which are falling onto galaxies and whose nature 
probably is cosmological (Blitz et al. 1998; L\'opez-Corredoira, 
Beckman, \& Casuso 1999). 

The structural and dynamical properties of disc galaxies as well 
as the correlations amongst them should be
self-consistently explained by theories of galaxy formation and evolution.
Among the local properties to be explained are the nature of the
exponential surface brightness or density distribution in discs, the
factors that determine the observed wide range of surface brightnesses 
(SB), the shapes of the rotation curves (typically flat) and their 
correlations with the SB. Among the global
structural and dynamical correlations of disc galaxies, the
Tully-Fisher relation (TFR) and the rotation velocity-radius
relation are of great interest. Understanding of the origin of
these observational relations ---particularly the TFR and its scatter---
is an outstanding problem of extragalactic astronomy. 
We would like to know whether the slope and zero-point of
the TFR are produced by evolutionary and/or SF processes (e.g. Silk
1997; Elizondo et al. 1998), or if they are direct imprints from a fundamental
scaling law of cosmic structures (e.g. Faber 1982; Frenk et al.
1988; Cole et al. 1994; MMW98; AFH98; Navarro \& Steinmetz 1998). 
Likewise we are interested in understanding why the TFRs for high 
and low SB galaxies are the same, and why the scatter of this relation 
is small. 

The purpose of this paper is to 
use an unified scenario of galaxy formation and evolution in 
the cosmological context in order to explain the items mentioned 
above. In a forthcoming 
paper, results concerning the luminosity properties, bulge formation, 
and the correlations which define the Hubble sequence will be 
discussed (Avila-Reese \& Firmani 2000). Since our aim is 
to study general trends, we shall 
only use one representative CDM model. It has the
following parameters: the matter density: $\Omega _m=0.35,$ the 
vacuum density: $\Omega_\Lambda =0.65,$ the Hubble
constant in unities of 100 kms$^{-1}$Mpc$^{-1}$: h$=0.65$, the 
amplitude of perturbations on a 8 Mpch$^{-1}$ scale:
$\sigma _8=1$. Hereafter, we shall refer to this model as the
$\Lambda $CDM$_{0.35}$ model.

In $\S 2$ we briefly describe the methods used to generate the MAHs 
and to calculate the virialization of the DM haloes. Results 
concerning the structural and dynamical properties of the DM haloes
are presented. The method used 
to build discs within DM haloes and to
follow their evolution, including SF and hydrodynamics, is presented
in $\S 3$. In $\S $4, local galaxy properties such as the surface
density, the rotation curve, and the decomposition of the rotation
curve are presented. Section 5 is devoted to the TFR, its scatter, and
the correlations between gas fraction vs. SB and $V_{\max}$
vs. disc size. We interpret why low and high SB galaxies 
lie on the same location in the TFR, and why
the residuals of the TFR are nearly independent of the 
SB. In \S 6 we explore how the introduction of a shallow 
core in the halo influences on the models. A summary and our 
conclusions are given in $\S 7$.
 
\section{Dark matter haloes}
 
\subsection{The method}

According to the hierarchical clustering scenario, the DM 
haloes are the environment in which luminous galaxies 
form and evolve. Therefore, some features of luminous 
galaxies depend on the evolution and properties of their 
surrounding dark haloes. In this subsection we present the method
we used to calculate the formation, evolution and structure
of the DM haloes. 

We calculate the gravitational collapse and virialization of 
DM haloes beginning from a primordial density fluctuation field. 
This was done in AFH98 and the reader is referred to it for 
details (see also Avila-Reese 1998). We assume a
Gaussian fluctuation field where all the statistical properties depend
only on the power spectrum of fluctuations given by the chosen cosmological
model. The conditional probability for Gaussian random fields is used to
calculate the mass distribution of haloes at time $t_{i+1}$ that will be 
contained in a halo of mass $M_i$ at a later time $t_i$ (Bower 1991; Bond et
al. 1991). Starting from a present-day mass and its cumulative 
density contrast, we apply recurrently this mass distribution 
through Monte Carlo simulations in order to construct different realizations
of the MAHs of ``main progenitors'' (Lacey \& Cole 1993).
Main progenitor in our case is the most massive subunit of the 
distribution, at each time, and not the larger mass between the chosen
mass $M_{i+1}$ and its complement $M'_i$, where $M'_i=M_i-M_{i+1}$.

To calculate the virialized profile of the dark 
halo that emerges from a MAH, we use a method that expands upon 
the secondary infall model (e.g. Gunn 1977;
Zaroubi \& Hoffman 1993) by allowing non-radial orbits and arbitrary
initial conditions (MAH in our case). Under the assumption of 
spherical symmetry, an initial density fluctuation ---given by the MAH--- 
can be described as a set of concentric
mass shells that sequentially attain their first maximum expansion 
radius $r_0$. After this maximum, the shell at $r_0$ which contains
mass $m_0(r_0)$, separates from the
expansion of the universe, falls toward the center due the
gravitational field of the internal mass, and evolves through
a radially oscillatory movement. Each oscilation defines an apapsis
radius $r_a$ which is function of $r_0$. The gravitational field
at radius $r$ is given by $m_T(r)$, the sum of the mass shells 
with $r_a\leq r$ that permanently oscillate inside $r$, $m_P(r)$, and 
of the mass shells with $r_a\geq r$ that only momentarily
fall inside $r$, $m_M(r)$. The estimate of $m_M(r)$ is 
obtained from the probability
\begin{displaymath}
P(r)\propto \int_0^r\frac{d\eta }{v\left( \eta \right)} 
\end{displaymath}
to find a shell inside a radius $r$($<r_a$), where $v\left( r\right)$ 
is the radial velocity of the shell. This velocity is: 
\begin{displaymath}
v^2\left( r\right) =2\left[ E-G\int_0^r\frac{m_T\left( \eta \right) }{\eta ^2%
  }d\eta -\frac{j^2}{2r^2}\right],
\end{displaymath}
where $G$ is the gravitational constant, $E$ is the total 
energy of the shell, and $j$ is the typical
angular momentum per unit mass of a shell mass element due to the
thermal motion which will be taken constant in time. The condition
$v=0$ defines the apapsis $r_a$ and a value $r_{p}$
which determines the maximum penetration of the shell towards the
center. We express $j$ through the parameter 
$e_0\equiv \left( \frac{r_{p}}{r_{a}}\right) _0$ and, 
although it is defined for
the first $r_{a}$ and $r_{p}$ of every shell, through $e_0$ we
are parametrically taking into account the thermal energy that could
be produced by the mergers of substructures and tidal forces at all
times. This parameter is calibrated to the results of N-body cosmological
simulations ($e_0 \approx 0.1-0.3$). The sequential 
aggregation of new shells, combined with their
motion toward the center, introduces new contributions to the
gravitational field which acts on the underlying shells. This
non-conservative spherical gravitational field changes $E$, and
consequently $r_a$, $r_p$, and $e$ of each shell. The contraction of $r_a$
leads to an asymptotic value which is identified as the
current virialized radius $r_v.$ The change of $E$ may be estimated by
assuming an adiabatic invariant for the orbital motion. A 
simple iterative numerical method allows us to calculate the
solution, i.e. the structure profile $m_T(r,t)$ at each time step.

The present-day mass used to initiate the Monte Carlo simulations 
will be called the nominal mass $M_{\rm nom}$. Our results show
that the outer mass shells within $M_{\rm nom}$ are not virialized. 
The mass shells that have been virialized are
roughly within the virial radius, $r_v$, where the mean
over-density drops below the critical value, $\Delta _c$,
given by the spherical collapse model; for the cosmological model used here 
$\Delta _c(z=0)=334$ (e.g. Bryan \& Norman 1998). Analysis of
numerical simulations show that at radii smaller than $r_v$,
mater is indeed close to virial equilibrium (Cole \& Lacey 1996; 
Eke, Navarro \& Frenk 1998). At radii between $r_v$ and
2$r_v$, matter is still falling onto the halo, while, at larger radii, 
matter is expanding with the universe. The mass contained
within $r_v$ is the virial mass $M_v$ which, depending upon the MAH, 
is 0.7-0.9 times $M_{\rm nom}$ (see also Kull 1999).

\subsection{Structure of the haloes: diversity}

In AFH98 it was shown that, due to the statistical
nature of the MAHs, a collection of {\it different }DM virialized
configurations are produced, the most typical configurations 
being reasonably well described by the density profile proposed by
Navarro, Frenk \& White (1996, 1997; hereafter NFW). 
Some features of the galaxies are related
to the dispersion about the virialized structures. Therefore, its
inclusion in the galaxy models is important. In the present work,
we apply a more accurate statistical treatment of the
results than in AFH98. Instead of initially selecting some relevant special
cases, we generate, for a given mass and through Monte Carlo
simulations, a catalog of objects with different MAHs (and
eventually different spin parameters $\lambda $ taken from a log-normal
distribution; see $\S 3$). Although most of the results of the statistical
analysis of this catalog agree with AFH98, we felt it was important to
perform this new procedure in order to obtain an accurate
estimate of the dispersion in the mass-velocity relation.



In Figs. 1a and 2a, a sample of twenty MAHs 
($M_{\rm nom}=5\times10^{11}M_{\odot }$)
and the circular velocity profiles of the haloes formed from them, are
plotted as functions of the collapse redshift, $z_c$. 
The average MAH and two representative cases, a fast
early collapse (L: low accretion rate at $z\approx 0$) and an extended 
collapse (H: high accretion rate at $z\approx 0$), are shown
in Fig. 1b. The circular velocity profiles corresponding to these
particular MAHs are plotted in Fig. 2b. These two MAHs were
chosen in such a way that roughly 95 per cent of all the
trajectories lie between of them. For a given mass, the average MAH 
is calculated as the mean of a large sample of trajectories. 

We find that a DM halo formed from an early fast MAH is
more concentrated than a halo produced by a gentle and extended MAH.
Nevertheless, as seen in Fig. 1a, the MAHs are diverse
and difficult to classify in uniparametrical, or even
biparametrical, sequences. That is why we have preferred to generate a
a catalog of MAHs for each given $M_{\rm nom}$ and to study the 
statistics of the DM haloes calculated from these MAHs {\it a posteriori}.

Some of the MAH trajectories show pronounced jumps, 
which indicates the occurrence of major mergers. Halo major merging
may imply disc major merging that probably destroys the discs. The 
fraction of MAHs that show evidence of at least
one major merger is between 20 and 30 per cent. When we use 
the results from our catalogs to calculate the TFR scatter, we keep 
all the MAHs since we want to obtain an upper limit of this scatter. 

\subsection{Comparison with N-body simulations and the $M_v-V_{\max}$ 
relation}

The density profiles of a large fraction of the calculated haloes are 
roughly described by the NFW profile. However, as it is seen in
Fig. 2b, the haloes present a diversity of structures. In 
Avila-Reese et al. (1999) the density profiles of the haloes
obtained with our method were compared with those of thousands
of haloes from a cosmological N-body
simulation. The statistical agreement found was rather good, 
in particular for the isolated haloes.
If we compare, for example, the statistical distribution 
of the outer density 
profile slope ($\beta$) for our haloes with those from the N-body 
simulations, we find they are very similar (see Fig. 4 in Avila-
Reese et al. 1999). Haloes in the N-body simulations present
a slightly broader distribution of $\beta$ than our haloes.

It is also important to mention that the MAHs of our haloes
for a given mass agree rather well with the mass evolution 
measured for the haloes in a cosmological N-body simulation (see Fig. 6
in Gottlober, Klypin \& Kravtsov 1999).

As in previous works, we find that the mass $M_v$
and the halo maximum circular velocity $V_{\max}$ obey a 
$M_v\propto V_{\max}^\alpha $ relation (in the 
$M_v\sim 4\times10^{10}M_{\odot }-4\times10^{12}M_{\odot }$ 
range and for the $\Lambda $CDM$_{0.35}$ model, $\alpha \approx 3.2$ 
and 3.3 from our results and from results of N-body simulations,
respectively). This relation is imprinted by the power spectrum 
of fluctuations and the MAH of the protohalo (AFH98; Avila-Reese 
et al. 1999). 

Owing to the statistical nature of the calculated MAHs, a scatter 
in the $M_v-V_{\max}$ relation is expected. A first attempt to 
estimate such a scatter was done by
Eiseinstein \& Loeb (1996) who used Monte Carlo simulations to
generate the MAHs, and the
simple spherical-collapse model to calculate the halo circular
velocities. AFH98 have estimated this scatter making use of their 
semi-numerical method for calculating the virialization process of 
the DM haloes. For the SCDM model both methods lead to 
similar results. Here, for the $\Lambda $CDM$_{0.35}$ model and in
the range of $M_v\sim 4\times10^{10}M_{\odot }-4\times10^{12}M_{\odot }$, 
we obtain fractional standard deviations
in the velocity, $\sigma _ V/<V_{\max}>$, from $\approx 0.10$ to $\approx
0.07$, respectively (see Table 1). The deviations estimated from the
cosmological N-body simulations are in rough agreement with 
these values, being only slightly larger than our results.
(see Fig. 10 in Avila-Reese et al. 1999). The deviations in velocity may
be translated into logarithmic standard deviations in mass:
$\Delta $log$M_v=\alpha $log$(1+\sigma _V/<V_{\max}>)$ (multiplying $\Delta$
log$M_v$ by 2.5, the standard deviation in mass can be expressed magnitudes). 
In column (4) of Table 1, the mass scatter of the $M_v-V_{\max}$ relation 
--- expressed in magnitudes --- is given for several masses. 


\section{Disc build up and galaxy evolution}

During or after DM reaches virial equilibrium, the baryon matter 
dissipates energy radiatively and falls to the
bottom of the gravitational potential well. If the DM halo has some
angular momentum, then a disc in centrifugal equilibrium forms at the centre
of the halo. In this section we describe the methods used to 
calculate the formation and evolution of discs. The main structural 
properties of the disc will depend on
the formation history of the halo, its structure, and on the
amount of infalling gas and its angular momentum. 

Analytical models of discs in centrifugal equilibrium
surrounded by DM haloes have been used to study several galaxy and galaxy
population features (e.g. Fall \& Efstathiou 1980; van der Kruit 1987;
Dalcanton et al. 1997; MMW98; van den Bosch 1988). The results obtained
with these models encourage us to study in more detail the extended
collapse scenario. Our contributions are: First, we were able to 
build up the disc sequentially
within an evolving DM halo. Second, we calculated the disc thickness 
and the multiple galaxy components in a 3-D gravitational potential. 
Third, we calculated the SF, disc hydrodynamics and ensuing galaxy evolution. 

The outline of our {\it evolution} models is:

{} (1) We consider that baryon matter has the same 
distribution of mass and angular momentum as DM till the 
accreted spherical (baryon+DM) mass shell virializes. 
We assume that each mass shell has a
solid body rotation, in agreement with the Zel'dovich
approximation, and that the rotation axis of the shells are
aligned. We did not assume that the whole halo is a solid
rotator (e.g. Dalcanton et al. 1997).  

{} (2) Once a given mass shell virializes, a 
fraction $f_d$ of its mass is transferred to a disc in centrifugal 
equilibrium. Since for galaxy haloes the 
time-scale of gas cooling is generally smaller than the
dynamical time-scale (c.f., Silk 1977; Rees \& Ostriker 1977; White \& Rees
1978; Ryden \& Gunn 1987) we assume the gas falls from the maximum
expansion radius of the current shell to the halo centre
in a time equal to the shell virialization time. The radial
distribution of the infalling gas is calculated by equating 
its specific angular momentum (the same of the DM component)
to that of its final circular orbit in centrifugal equilibrium. 
The specific angular momentum $j_{sh}$ acquired by
each collapsing mass shell during the linear regime is estimated under the
assumption of a constant spin parameter $\lambda $ ($\equiv 
\frac{J\left| E\right| ^{1/2}}{GM^{5/2}}$): 
\begin{equation}
j_{sh}(t)=\frac{dJ(t)}{dM(t)}=\frac{{GM(t)^{5/2}\lambda }}{\left|
E(t)\right| ^{1/2}}\Bigl(\frac 52\frac 1{M(t)}+\frac{d\left| E(t)\right| }
{2dM(t)}\Bigr),
\end{equation}
where $J$, $M$, and $E$ are the current total angular momentum, mass, 
and energy of the halo at time $t$.

{} (3) The gravitational drag on the dark halo produced by the collapse
of each baryon mass shell is calculated through the adiabatic 
invariant formalism (e.g. Flores et al. 1993).

{} (4) We consider that the {\it local} SF and internal 
hydrodynamics of the disc are regulated by a balance between 
the kinetic energy injected by SNe and gas accretion,
and the energy dissipated by the turbulent interstellar 
medium (Firmani et al. 1996, 1997). The star formation is 
turned on at radius $r$ when the local Toomre gravitational
instability parameter for the gas disc, 
$Q_g(r)\equiv \frac{v_g(r)\kappa(r) }{\pi G\Sigma _g(r)}$, 
falls below a given threshold; $\kappa (r)$, $v_g (r)$, 
and $\Sigma _g (r)$ are the epicyclic frequency, the 
gas velocity dispersion, and the gas surface density at 
radius $r$, respectively. Thus, 
the SF is controlled by a feedback mechanism such
that, when a gas disc column is overheated by the SF activity, SF 
is inhibited and the disc column dissipates the excess energy 
to lower $v_g$ back to the value
determined from the Toomre criterion threshold. Numerical simulations
(Sellwood \& Carlberg 1984; Carlberg 1985; Gunn 1987) and observational
estimates (e.g. Skillman 1987; Kennicutt 1989) suggest thresholds of the
order of 2 instead of 1, as was analytically obtained for a
thin disc (Toomre 1964). This difference is attributed to collective
phenomena which are difficult
to account for in the analytical studies. In our models, the value of this
threshold controls the thicknesses of the gas and stellar discs; the SF
rate actually is rather insensitive to the value of $Q_g$. When a
value of 2 is used, for a model of the Galaxy, we obtain gas and stellar disc
thicknesses compatible with those of the solar neighborhood. Thus, we fix
$Q_g=2$. The gas loss from stars is also
included. The gravitational dynamics of the evolving star and gas discs, and
the DM halo are treated in detail. A Salpeter initial mass function is
assumed. Analytical fits to simple stellar population
models are used in order to calculate the luminosity in the $B$ band (see
for more details Firmani \& Tutukov 1994; Firmani et al. 1996).

Note that in our self-regulating SF mechanism the feedback
happens only within the disc and not at the level of the whole halo.
This is justifiable since the turbulent interstellar medium is a very 
dissipative system (e.g., Avila-Reese \& V\'azquez-Semadeni 2000).
Gas and energy outflows are confined within a region close to the disc. 
The thick gaseous disc and the global magnetic field are efficient shields
that prevent any outflow toward the halo on large scales (e.g.
Mac Low, McCray \& Norman 1989; Slavin \& Cox 1992; Franco et al. 
1995). The lack of observational evidence for significant amounts 
of hot gas in the haloes of disc galaxies
confirms these studies and suggests that the halo-disc connection is
not enough to self-regulate and drive the disc SF. Some observational
evidence shows that extragalactic gas clouds reach the disc 
in free fall (see Blitz et al. 1998). 

According to the scheme described above, time by time 
and at each radius, the growing disc is characterized 
by the infall rate of fresh gas by unit of area, 
$\dot {\Sigma _g}(r,t)$, the gas and stellar disk surface 
density profiles, ${\Sigma _g}(r,t)$ and ${\Sigma _s}(r,t)$, 
the total rotation curve (including the growing DM halo 
component), $V_r(r,t)$, and the SF rate $\dot {\Sigma} _s(r,t)$ 
determined by the energy balance in the vertical 
gaseous disk and by a Toomre criterion. It is interesting to 
point out that a consequence of our physical model of SF
is that $\dot {\Sigma} _s$ exhibits a Schmidt law 
with index $\approx 2$ at all radii and during almost all 
the evolution (Firmani et al. 1996; Avila-Reese \& Firmani 2000).
It should be mentioned that the global SF rate
efficiency in our models does not depend on the mass or 
$V_{\max}$ of the galaxy (see Avila-Reese \& Firmani 2000). The 
global SF rate efficiency is mainly a function of ${\Sigma _g}$ 
(determined by $\lambda$) and $\dot {\Sigma} _g$ (determined by the MAH).    

Our goal is to generate a catalog of models for each given present-day nominal
mass, $M_{\rm nom}$ (see \S 2). The key initial factors for a model of a given
mass are (i) the MAH (it determines the structure of the halo and the rate at
which the gas is accreted onto the disc), (ii) the spin
parameter $\lambda $ (strongly influences the size and surface density
of the disc), and (iii) the effective baryon fraction  $f_d$
that is incorporated into the disc. The MAHs (see Fig. 1) and
the halo evolution are calculated as was described in $\S 2$. The spin
parameter $\lambda $ is taken to be constant in time and is chosen 
from a log-normal distribution through Monte Carlo simulations. 
The median and the dispersion of the log-normal distribution
we use are 0.05 and 0.5, respectively, in agreement with results of
several theoretical and numerical studies (see MMW98 and the references
therein). Regarding item (iii), in a first approximation one might take the
baryon-to-dark matter ratio of the protogalaxies equal to that  of the 
universe, $f_b=\Omega _b/\Omega _m$. According to the current primordial
abundance determinations of the baryon density, $\Omega _bh^2\approx
0.006-0.013$ (e.g. Fukugita, Hogan \& Peebles 1998), the baryon-to-dark
matter fraction of the universe is $f_b \approx 0.04-0.09$ for the
cosmological model used in this paper ($\Omega _m=0.35$, 
$\Omega _{\Lambda}=0.65$, $h=0.65$). However, it is probable 
that not all the halo baryon fraction  ends up in the galaxy central disc;
some gas fraction may remain in the form of hot halo gas or may also be
transformed into stars within some small halo substructures. Here we shall
assume that the disc contains 5 per cent ($f_d=0.05$) of the total halo mass (see
also Dalcanton et al. 1997; MMW98). We are aware that this is a
free parameter in our models and when necessary, we shall explore here models
for other $f_d$ values.


\section{Local properties of disc galaxies}

In this section we present the stellar surface density
and rotation velocity profiles, and the rotation
curve decompositions for our model galaxies.
Most of the results in this section refer to models 
with the H, average, and L MAHs (see $\S 2$), and with 
$\lambda =0.03, 0.05,$ and 0.08. 

\subsection{Stellar surface density}

The stellar surface density (SSD) profiles of the rotationally supported discs,
sequentially formed within the evolving CDM haloes, are nearly exponential
over several scale lengths. Fig. 3 shows the SSD profiles for models of 
$M_{\rm nom}=5\times10^{11}M_{\odot}$ with several representative values
of the MAH (panel a) and the spin parameter $\lambda $ (panel b). It
is worth remarking that, in contrast to the models of Dalcanton et al. (1997; 
see also Fall \& Efstathiou 1980), we have not assumed the whole 
protogalaxy as a solid body rotator (see $\S 3$).
Therefore, the exponentiality of the stellar disc profile is a
prediction of our model where the crucial assumption made regarding
disc formation is the time constancy of $\lambda $. Analysis of 
results of N-body simulations regarding the evolution of the spin 
parameter of the haloes seem to confirm this assumption (Gottlober, 
private communication). We find the same
general trends as in Dalcanton et al. (1997) and  Jimenez at al. (1998): 
the central SSD depends strongly upon $\lambda$, and less upon the
mass. We also find that the SSD has some dependence upon the MAH (Fig. 3a). 
We conclude that the low SB galaxies may be mainly galaxies with high
angular momenta and/or low masses. The value $\lambda \approx 0.05$ 
probably introduces a natural separation between
high and low SB galaxies.

The characteristic radii and SB of our disk galaxy models
are realistic (see also Avila-Reese \& Firmani 2000). The 
assumptions of detailed angular momentum conservation and 
rotation axis aligment of the shells were crucial for 
these predictions. According to numerical 
simulations of galaxy formation, the former assumption
seems to fail (e.g., Navarro\& Steinmetz 1997), posing
a difficulty for the hierarchical formation scenario. Possible
intermediate astrophysical processes might solve this difficulty 
(e.g., Weyl, Eke \& Efstathiou 1998). Regarding the latter
assumption, according to preliminar results of analysis of 
CDM cosmological N-body simulations, it seems to be reasonable,  
at least for the present-day isolated haloes (Avila-Reese, Klypin, 
Firmani, \& Kravtsov, in preparation), although
deeper studies are necessary, in particular for the inner regions
of the halos whose collapse occured at early epochs.


\subsection{Rotation curves}

The shapes of the rotation curves depend 
on $\lambda $, on the MAH that determines the DM halo
structure, and on $f_d$. In Fig. 4, we plot rotation curves
corresponding to models of $M_{\rm nom}=5\times10^{11}M_{\odot }$ 
for a variety of MAHs and spin parameters. From Fig. 4,
the main tendencies appear clearly: the rotation curve shapes are
more peaked as $\lambda $ is smaller or the MAH is more active at early
epochs (the DM halo is more concentrated).
Due to the correlation between SSD and $\lambda$, a correlation between
the SSD (SB) and the rotation curve shape is natural
in our models: high SSD galaxies typically
present more peaked rotation curves than low SSD
galaxies. Observational studies seem to find a similar trend with
SB (e.g. Casertano \& van Gorkom 1991; Verheijen 1997).
Although less significant, observations also show a dependence 
of the rotation curve shape on luminosity (Persic \& Salucci 1988; Casertano
\& van Gorkom 1991; Persic, Salucci, \& Stel 1996). Our models tend to
confirm this dependence (see Avila-Reese \& Firmani 2000). Finally, for high
values of $f_d$, the shapes of our rotation curves are more peaked (Fig. 5). 


As it was previously  pointed out by MMW98, we find that the minimum
spin parameter, $\lambda _{\min}$, for which the rotation curves are
still realistic (nearly flat), should be increased as $f_d$
increases. For instance, an eye inspection of panels (a) and (d) of
Fig. 5 suggests that when $f_d=0.05$ the $\lambda_{\min}$ can be
as low as 0.03, while $\lambda_{\min}$ probably should be
increased (to $\sim 0.04$) when $f_d=0.08$. Dalcanton et al. 
(1997) and MMW98 proposed that models with 
$\lambda <\lambda _{\min}$ are gravitationally unstable. For example, 
according to the predicted distribution of $\lambda$, this means that if 
 $\lambda _{\min}\approx
0.04$, then roughly one third of the galaxies have unstable discs (S0's,
ellipticals?), a fact that disagrees with the observations in the
local universe. This problem is enhanced if some angular momentum
is transfered during the gas collapse,  as is seen in the N-body+hydrodynamics 
simulations. This difficulty could be attenuated if the DM haloes 
would have inner density profiles shallower than $r^{-1}$ (see $\S 6$).


\subsection{Rotation curve decompositions}

In Fig. 5 we show the decompositions of the rotation
curves into disc (stars+gas) and halo (contracted by the baryon
collapse) components for models of $M_{\rm nom}=5\times10^{11}M_{\odot }$ 
and different values of $\lambda $ and $f_d$. As is seen in 
Fig. 5, the maximum of the disc rotation
velocity is attained at $\sim 2.2R_s$. The disc/halo
decompositions of the observed rotation curves do not have a single solution, 
and an additional constraint is required in order to reproduce them. 
The ``minimum halo'' (or
``maximum disc'') solution (c.f., Carignan \& Freeman 1985; Sancisi \& van
Albada 1987) has been the commonly adopted restriction. In this case, 
the ratio between the disc and total maximum rotation velocities,
$V_d/V_t$, is about $0.85$ (e.g. Sackett 1997). Nevertheless, as some 
observational works have pointed out, this solution seems to present some
shortcomings or at least is not applicable to all galaxies 
(Persic \& Salucci 1988; see Navarro 1998 for more references). From an
analysis of stellar velocity dispersions Bottema (1993, 1997), concluded that
$V_d/V_t$ is about 0.63 for high SB galaxies, and even less for low SB
galaxies. 

In Fig. 6, $V_d/V_t$ is 
plotted as a function of the SSD for 20 models from our catalog
corresponding to $M_{\rm nom}=5\times10^{11}M_{\odot }$ and $f_d=0.05$
(filled circles). The disc contribution increases as the
galaxy SSD is greater. The models  corresponding to high SSD galaxies 
($\Sigma _{s,0}\approx 200-2000 M_{\odot }/pc^2$, 
where $\Sigma _{s,0}$ is the central SSD) have
on average $V_d/V_t\sim 0.70$. This is the approximate value of 
the average galaxy model with $\lambda =0.05$ for which the DM dominates at 
essentially any radius (Fig. 5). Some theoretical arguments (e.g.
Athanassoula et al. 1987; Debattista \& Sellwood 1998) and
observational studies (Verheijen 1997; Corsini et al. 1998)
suggest that in high SB galaxies the disc
component should dominate to some extent at the most inner radii. 

In
our models, the dominance of the halo component over the disc component
is mainly due to the steep inner density profile ($\propto r^{-1})$ of
the original DM halo. Therefore, the inferred rotation curve
decompositions of observed galaxies seem to point out that the halo core 
should be shallower than $r^{-1}$. More direct evidence for shallow halo
cores comes from the rotation curves of dwarf and low SB galaxies (Moore 1994; 
Flores \& Primack 1994; Burkert 1995). In $\S 6 $ we shall explore
models where shallow cores are artificially introduced to our
DM haloes.


\section{Structural and dynamical correlations of disc galaxies}

In this section we present the $M_s-V_{\max}$ (TF) and $R_s-V_{\max}$
relations calculated from our catalogs of galaxy models. The 
scatter of the $M_s-V_{\max}$ relation is studied in detail. We
analise the correlation among the residuals of the above
relations and we explain why low and high SB galaxies have the same TFR.

\subsection{Infrared Tully-Fisher relations}

Disc galaxies present a strong correlation between their
luminosities $L_i$ ($i$ is the spectral band) and their maximum 
rotation velocities $V_{\max}$, commonly known as the TFR 
(Tully \& Fisher 1977). In the infrared bands ($i=I,H,K,...)$, 
this relation is given by:

\begin{equation}
L_i=A_iV_{\max}^{\rm m_i} 
\end{equation}
where $A_i$ is related to the so called zero-point, and $3\lesssim 
m_i < 4$ according to several observational studies. Since 
$L_i\propto M_s,$ where $M_s$
is the disc stellar mass, eq. (2) may be interpreted as a relation between 
$M_s$ and $V_{\max}$. In Fig. 7, where $M_s$ is plotted as a function of
 $V_{\max}$, we present our results from the
catalog constructed by Monte Carlo simulations with $f_d=0.05$ 
(see \S 2 and \S 3). The error bars represent the standard deviation
calculated by adopting a normal distribution for the deviations of
the velocity for a given mass; see also Table 1. In the range of masses
considered here ($M_s\approx 10^9-10^{11}M_{\odot }),$
the slope of the $M_s-V_{\max}$ relation is approximately 3.4; this slope
is slightly larger than the slope of the mass-velocity relation of the
cosmological DM haloes ($\S 2$). 

In Fig. 7 we have also included observational data. In the 
$I$ band we used the TFRs given by Giovanelli et al. (1997) and by 
Willick et al. (1995; they used the data published by Han,
Mould and collaborators, see the references therein). In these 
studies the line widths were corrected for non-circular motions.
In the $H$ band we used the TFRs given by Gavazzi (1993) and by
Pelletier \& Willner (1993). Although the $H-$magnitudes of most of the
galaxies reported in Gavazzi (1993) were obtained through aperture photometry,
he used the total $H-$magnitudes estimated
with an extrapolation technique. We corrected the TFR given 
by Gavazzi for non-circular motions (see AFH98 for details).
In the case of Pelletier \& Willner, we used their TFR calculated 
with the total $H-$magnitudes obtained with an infrared array and the 
line widths corrected for non-circular motions.  Regarding 
the determination
of $V_{\max}$ in the observational studies we plot in Fig. 7, most 
of the data were obtained from HI line-width measures. As 
Verheijen (1997) have shown, the differences in the slope of the TFR 
calculated with single dish and detailed synthesis data are small;
if any, the slope is slightly shallower when using the detailed synthesis
data.

 In earlier works about
the near infrared (particularly $H-$band) TFR with aperture photometry, 
the slope obtained was $\sim 4$; however, as pointed out by Pierce 
\& Tully (1988) and Bernstein et al. (1994), the use of aperture 
magnitudes results in an artificially large slope to the TFR. This
is why in studies where $CCD-$photometry is used the slope of the 
TFR in infrared bands resulted shallower than 4\footnote{After the 
complexion of this paper a study by Tully \& Pierce (1999) appeared 
where the authors carrefully re-evaluate observational
data in order to accurately determine template TFRs in different 
bands. They conclude that there appears to be convergence in the 
infrared towards a TFR's slope of $3.4\pm 0.1$ (see also Rohtberg et al. 1999)} 
(e.g., Pierce \& Tully 1988; Pelletier
\& Willner 1993; Bernstein et al. 1994; Verheijen 1997). The last 
author, for a sample of galaxies in the Ursa Major cluster obtained
a $K'-$band TFR with slope $\approx 3.3$ when he used the complete 
sample of 41 galaxies and with slope $\approx 4$ when he used
only 15 galaxies out from the sample, selected to be unperturbed galaxies
of late type Sb-Sd and without prominent bars.
      
In order to transform luminosities into stellar masses, 
a mass-to-luminosity ratio, $\Upsilon $, should be adopted. 
For the $I-$band observations, we assumed 
$\Upsilon _I=1.8\left( \frac{M_s}{5\times 10^{10}M_{\odot }}\right) ^{0.07}h$ 
(see AFH98). This ratio is close to the one suggested by 
MMW98 ($\Upsilon _I=1.7h$ ) on the basis of the 
$\Upsilon _B$ that Bottema (1997) inferred from disc dynamics. For the
$H-$band observations $\Upsilon _H=0.55$ was
used and $h=0.65$ was assumed. This mass-to-luminosity ratio is obtained
from direct observational estimates in the solar neighborhood (Thronson \&
Greenhouse 1988; see details in AFH98). 

In Fig. 7, the model results are slightly shifted to the high 
velocities with respect to the Giovanelli et al. (1997) and Gavazzi (1993) data, 
while they agree rather well with the Han-Mould (quoted by Willick et al.
1995) and Pelletier \& Willner (1993) data.
The assumed $f_d$ does not significantly change these 
results because our models typically shift along the main relation for different
values of $f_d$ (see below). In order to reduce the 
uncertainties due to $\Upsilon $ we have used measured TFRs in two
bands ($I$ and $H$) with $\Upsilon $ independently estimated. 
 The tendency of the models to have a larger $V_{\max}$ 
for a given $M_s$ may again be showing that the CDM haloes are too cuspy.
When a shallow core is introduced in the haloes (see \S 6), 
the resulting $V_{\max}$ of the galaxies are
slightly smaller (empty circles in Fig. 7). Despite the 
uncertainties, we conclude  that the agreement 
of the slope and the zero-point of the TFR  between observations and our 
theoretical results is reasonable well. Note that in the theoretical 
calculations, both the cosmological framework and the SF process 
were taken into account. 

Our results show that the TFR is mainly a product of the mass-velocity 
relation of the DM haloes and, as stated in $\S 2$, this is
determined by the cosmological initial conditions (e.g. Frenk et al. 1988, 
Cole et al. 1994, MMW98, 
AFH98, Steinmetz \& Navarro 1998). The slope of the TFR is linked to the 
shape of the power spectrum at galaxy scales and the MAH of the halo. 
For most of the CDM models, the power spectrum shape at galaxy scales and the 
MAHs are almost the same, so the slope of the TFR is expected to be a
generic feature of the CDM cosmogony (Firmani \& Avila-Reese
1999). Concerning the zero-point, from the point of view of the 
cosmological model it depends on the amplitude of the power 
spectrum at galaxy scales. Models with low amplitudes
at galaxy scales ---like the $\Lambda $CDM or open CDM models--- predict 
the zero-point of the TFR better than those with high amplitudes
(for example, the standard CDM model; see also Firmani \& Avila-Reese
1999; Jimenez \& Heavens 1999). 


The results presented in Fig. 7 were obtained assuming $f_d=0.05$.
We find that the $M_s-V_{\max}$ relation is rather insensitive
to changes in the disc mass fraction, particularly when this fraction
is such that the disc contribution to the rotation curve is relevant. 
In the right lower corner of Fig. 7 the solid
line represents the shifts that a typical galaxy model (average MAH, $\lambda
=0.05$) suffers when varying $f_d$. From right to left the dots correspond
to models with $f_d=0.03$, 0.05 and 0.08, respectively. For the
typical galaxy models, an increment in the parameter $f_d$
basically produces a shift along the $M_s-V_{\max}$
relation. This is due to the ``compensating'' action
of the gravitational pull exerted by the disc on the DM halo. 
This result also suggests that, if $f_d$ fluctuates or has a dependence upon 
mass, by effects of gas cooling or feedback, then 
the TFR is almost unaffected, at least in the range mentioned above.
When $f_d$ is very small, the disc contribution
to the rotation velocity is negligible,  and $V_{\max}$ is determined by 
the DM halo alone. In this case, the $M_s-V_{\max}$ relation becomes 
sensitive to $f_d$ because the stellar mass (luminosity) and the galaxy 
dynamics are de-coupled.

Based on the results presented above and as was pointed out 
in Firmani, Hern\'andez, \& Avila-Reese (1997), we find that 
the infrared TF relations offer a robust way to normalize the 
power spectrum of fluctuations at galaxy scales independently 
of uncertainties due to the disc mass fraction assumed.
This normalization favours power spectra corresponding to 
$COBE$-normalized low density CDM models ($\Lambda $CDM or 
open CDM models).

\subsection{Scatter of the Tully-Fisher relations}

The observational rms scatter of the TFR is
small. In the $I$-band Giovanelli et al. (1997) reported a total scatter of
$\sim 0.3$ mag for rotators of $\sim 180$ km/s; they found that the scatter
increases from fast to slow rotators. Willick
et al. (1995, 1996) and Mathewson and Ford (1994) estimated a total scatter of
0.38-0.43 mag and 0.44 mag, respectively. Bernstein et al. (1994) found a 
scatter of 0.23 mag. In the $H$-band Willick et al. (1996)
found a scatter of 0.47 mag. These last authors
concluded that the estimated {\it intrinsic} scatter of the infrared TFRs 
is not smaller than 0.3 mag.
Are the theoretical models able to predict the intrinsic 
rms scatter inferred from the 
observations? In our models, there are at least two sources 
of the scatter in the infrared TFR. One is related to the statistical nature 
of the MAHs. The dispersion in the MAHs determines
a dispersion in the maximum circular velocities of
the virialized structures (Eisenstein \& Loeb 1996; AFH98; see $\S 2$). The
second source is associated with the dispersion of the spin parameter $\lambda
$ (e.g. MMW98). As $\lambda $ decreases, the disc becomes 
more concentrated, and therefore  $V_{\max}$ increases. 

A first qualitative estimate of the scatter produced by the MAH and $\lambda $ 
distributions may be appreciated by the shifts from the 
$M_s-V_{\max}$ relation shown in the right lower corner of Fig. 7. 
The dotted and dashed lines correspond  to different  MAHs and 
$\lambda's$. Dispersions in both the MAH and $\lambda$
tend to shift the models along the $M_s-V_{\max}$ relation, in a similar
way as $f_d$. In columns (3) and (4) of Table 2, we present 
the total scatters in velocity and mass of the $M_s-V_{\max}$ 
relation obtained from our catalogs for three stellar disc
masses. The scatter in mass is expressed in magnitudes (see $\S
2.3$). There is a marginal agreement with the observations. Observational 
and theoretical uncertainties are large, so 
it is still premature to draw any conclusions relating the 
scatter of the TFR and the cosmological initial conditions
(but see Eiseinstein \& Loeb 1996). It should be noted that
we have maximized the scatter in our models by including both low and
high SB galaxies, and taking into account major mergers in the
calculation of the MAHs. The observational estimates
refer mostly to high SB late-type galaxies. On the other
hand, we have not included any source of scatter related 
to non-stationary SF or possible variations in $f_d$. Observational 
scatter due to non-stationary SF and $f_d$ would imply a larger 
scatter in our models, although Elizondo et 
al. (1998) have shown that, in this case, compensating effects may play
an important role in keeping, or even decreasing, the scatter of the TFR.  


We calculated a set of models with a constant $\lambda= 0.05$, in order to 
estimate the scatter of the $M_s-V_{\max}$ relation, due to variations
in the MAHs. 
In order to estimate the contribution to the scatter of the $M_s-V_{\max}$ 
relation due to $\lambda $, we calculated a set of models using the 
average MAH and taking the different $\lambda $ values from its log-normal
distribution. We find that the influence of $\lambda$ on the scatter of 
the $M_s-V_{\max}$ relation is smaller than that due to
the MAHs; the quadratic contributions of the scatter in $\lambda$
and in MAH to the total scatter, are roughly  25
and 75 per cent, respectively. 

\subsection{Residuals of the $M_s-V_{\max}$ and $M_s-R_s$ relations 
and the TFR of low SB galaxies}

In our models $\lambda$ strongly influences the SSD. Consequently,
the small contribution of $\lambda$ to the scatter of the $M_s-V_{\max}$ 
relation, implies that galaxies of different SB should have 
almost the same TFR. This agrees with the observations
that show that the TFRs of low and high SB galaxies are the same (e.g.
Zwaan et al. 1995; Tully \& Verheijen 1997; Verheijen 1997).

Recently, Courteau \& Rix (1998), using large catalogs of late-type,
high SB galaxies, have studied the correlations among the residuals 
of the TF and the luminosity-radius relations. They find 
that the slope of the correlation among the residuals, 
$\partial V_{2.2}/\partial R_s$, has a mean value of 
$-0.19\pm 0.05$ ($V_{2.2}\approx V_{\max}$ is the
value of the rotation velocity at $2.2R_s$). According to 
Courteau \& Rix (1998), this means that the TFR scatter  
correlates only slightly with disc size or SB. They interpret 
this as a evidence for large amounts of DM in the inner parts 
of late-type galaxies ($R<R_s$). If the DM is dominant, the 
disc component plays almost not role in setting $V_{\max}$. 
This conclusion, however, might not be consistent with observations 
that hint that shape of the rotation curve correlates with SB for
a given luminosity (Casertano \& van Gorkom 1991; Verheijen 1997).


Using our models we can explain this apparent observational inconsistency.
In order to interpret the observational results of Courteau \& 
Rix (1998) one must take into account the difference in the SF histories 
among  galaxies of different SB. From our models, in Fig. 8 we plot 
the deviations from the $M_s-V_{\max}$ relation, $\delta $log$V_{\max}$, vs. 
the deviations from the $M_s-R_s$ relation, $\delta $log$R_s$. The 
model galaxies were divided into
three groups depending on their central SSD:
very high SSD ($\Sigma_{s,0}>2000M_{\odot }/pc^2$), high SSD 
($200 M_{\odot}/pc^2<\Sigma_{s,0}<2000 M_{\odot}/pc^2$),
and low SSD ($\Sigma_{s,0}<200 M_{\odot }/pc^2$); we used 
black, gray, and empty circles to represent them. The slope of the
correlation among the residuals changes with the SSD of the galaxy.
The very high SSD models probably do not represent realistic
disc galaxies because their rotation curves are too peaked. These 
models could be subject to instabilities 
that destroy the disc (see $\S 4.0.2$). For the high SSD models the slope 
is approximately $-0.15$ in agreement with the results of Courteau \& 
Rix (1998). In order to compare the models with observations that 
include a wide range of SB's, we estimated the residuals 
$\delta $log$V_{\max}$ and $\delta $log$R_s$
using the data given in Verheijen (1997) and Tully et al. 
(1996) for galaxies of the Ursa Major cluster. Fig. 8b shows these
residuals where filled and empty circles represent high and low SB galaxies,
respectively. In spite of the small number of objects, the 
tendencies in this plot are similar to those
of our models. The low SB galaxies specifically demonstrate that
there is no extrapolation of the trend found by Courteau \& Rix (1998) to 
large radii (low SBs). 

From Fig. 8a, it is seen that for galaxies with high SSD (small 
$\delta $log$R_s$) the deviates from the $M_s-V_{\max}$ relation
decrease toward the low velocity side as the SSD decreases. However,
for low SSD galaxies (large $\delta $log$R_s$), this behavior is 
reversed. This result is due to the fact that not only
$V_{\max}$ decreases as the SSD decreases, but so does the
{\it stellar} disc mass, $M_s$. For this reason, for 
a {\it fixed} $M_s$ increasing the scale length $R_s$ (decreasing the SSD),
two regimes are present in Fig. 8a. When
the SSD is large ($\Sigma_{s,0}\grtsim 200 M_{\odot }/pc^2$), the 
disc contribution to the circular velocity becomes important and the
residuals of the $M_s-V_{\max}$ relation decrease with the residuals
of the $M_s-R_s$ relation ($V_{\max}$ decreases with $R_s$). But, when 
the SSD is small ($\Sigma_{s,0}\lesssim 200 M_{\odot }/pc^2$) the 
fraction of gas converted into stars is smaller than in discs 
with larger SSDs. Thus, in the last case $M_s$ should correspond to
a galaxy with a more 
massive halo than in the former case. Therefore, the 
residuals of the $M_s-V_{\max}$ relation increase with the residuals
of the $M_s-R_s$ relation ($V_{\max}$ increases with $R_s$). At the
same time, when the SSD is large the rotation curve is 
dominated by the disc and it results peaked, while when 
the SSD is small, the rotation curve is dominated by the halo 
contribution, it slowly increases, and presents a broad maximum.     

We conclude that the dependence of $M_s$ upon the SSD is responsible 
for the almost flat and non-monotonic slope of the correlation among the
residuals of the $M_s-V_{\max}$ and $M_s-R_s$ relations (Fig. 8a,b).
One expects that this slope becomes steeper and monotonic if the total 
(stars+gas) disc mass $M_{\rm tot}$ is used instead of $M_s$. Since 
$M_{\rm tot}$ does not depend upon the disc surface density, then 
as the disc surface density increases (i.e. the disc scale length 
$R_{\rm tot}$ decreases), $V_{\max}$ increases for all the SB's. 
The residuals from the 
$M_{\rm tot}-V_{\max}$ and $M_{\rm tot}-R_{\rm tot}$ relations, for the
same models from Fig. 8a are plotted in Fig. 8c. As can be seen
from Fig. 8c, if we assume that the disc stellar mass 
(luminosity) is exclusively proportional to the total disc mass
(e.g. Dalcanton et al. 1998; MMW98; AFH98; van den Bosch 1998), 
then we arrive to an incorrect result: the slope of the correlation 
among the residuals of the TF and luminosity-radius relations is 
steep and monotonic, or in other words, the deviations from the 
TFR is highly correlated with the SB. This contradicts the
observational inferences of Courteau \& Rix (1998), and that based 
upon the data from Verheijen (1997) and plotted in Fig. 8b. 

In our model galaxies, the stellar mass $M_s$ depends upon the
disc surface density, because the SF efficiency depends 
upon the disc surface density. The stellar mass in our models, also 
depends upon the gas accretion rate given by the MAH. In Fig.
9 we have plotted the fraction of gas ($f_g=M_{gas}/(M_{gas}+M_s))$ 
as a function of the central SB in the $B$ band, for our models
and for the data compiled by de Blok \& McGaugh (1997). The agreement 
between models and observations is rather good, suggesting that 
our SF histories are realistic. We conclude that
the disc surface density plays an important role in determining
the gas fraction, $f_g$, of disc galaxies.


The analysis presented in this subsection allows one to understand
why high and low SB galaxies have the same TFR, even though
the shapes of their rotation curves depend upon the SB. 
In the $M_s$ vs. $V_{\max}$ plot (Fig. 7), as the SB decreases,
the galaxies not only shift towards lower velocities, but also shift
towards a smaller stellar mass (luminosity). Thus, the deviations from
the $M_s-V_{\max}$ relation (TFR) are almost independent of the
disc size or SB. This effect also explains why the 
scatter about the $M_s-V_{\max}$ relation due to the
dispersion of the parameter $\lambda $ becomes so small (see $\S 5.2$).

\subsection{Disc sizes vs. maximum rotation velocities}

The distribution of disc galaxies in the $R_d-V_{\max}$ diagram is
shown in Fig. 10a. The data were taken from a catalog of late-type
galaxies in the $r-$band elaborated by Courteau (1996,1997) (circles), 
and from a sample of galaxies of the Ursa Major cluster in the 
$K'-$band elaborated by Verheijen (1997) (triangles). We transformed 
the surface brightness in the $r-$ and $K'-$bands to surface densities
using $\Upsilon _r=1.4$ and  $\Upsilon _{K'}=0.6$. We 
have tentatively divided the samples into two groups:
high SB ($\Sigma_{s,0}\grtsim 200 M_{\odot }/pc^2$, filled symbols) and 
low SB ($\Sigma_{s,0}\lesssim 200 M_{\odot }/pc^2$, empty symbols). 
The threshold we used for the division in Courteau's (1996,1997) sample was
$\mu_0=20.5$ R-mag/arcsec$^2$ (low SB galaxies are under-represented in
this sample). For the Verheijen (1997) sample the threshold we used was
18.5 $K'-$mag/arcsec$^2$. Despite the incompleteness of the samples
and the differences between them, Fig. 10a indicates
that the disc size correlates with the maximum rotation velocity,
particularly for families of similar SB. In Fig. 10b we
plot the results from our model catalogs corresponding to
three masses ($M_{\rm nom}=5\times10^{10}M_{\odot }$, $5\times10^{11}M_{\odot }$,
and $5\times10^{12}M_{\odot }$). As was noted in the previous subsection, 
the very high SB discs with $\Sigma_{s,0}>2000M_{\odot }/pc^2$ 
($\mu _0<16$ K'-mag/arcsec$^2$; black filled circles) probably are 
not realistic; they might be subject to gravitational instabilities
(these models have $\lambda <0.025$).    
The solid lines are the linear regressions for the normal and low SSD
models (gray and empty circles, respectively). These lines are
reproduced in panel (a) for comparison with the observations. The
agreement is reasonably good.

In the lower right corner of Fig. 10b, we show how galaxy models
($M_{\rm nom}=5\times10^{11}M_{\odot }$) shift in the $R_d-V_{\max}$
plane when the MAH, $\lambda $, and $f_d$ are varied. The largest 
shift is due to $\lambda $, which is why the $R_d-V_{\max}$ relation 
is well defined only for discs of similar SSD (SB).


\section{Effects of a shallow core}

From the observed rotation curves of some dwarf and low SB 
galaxies, Kravtsov et al. (1998, hereafter KKBP98) empirically
inferred an approximate self-similar density profile for the haloes of
these galaxies assuming that they are completely dominated by DM:
\begin{equation}
\rho (r)=\frac{\rho _0}{(r/r_0)^{\gamma }[1+(r/r_0)^{\alpha }]^{(\beta
-\gamma)/\alpha}},
\end{equation}
with $(\alpha,\beta,\gamma)=(2.0,3.0,0.2)$, where the particular value of 
$\gamma =0.2$ is weakly motivated; it should be considered just as
the evidence of a shallow core ($\gamma \approx 0$,  see also Burkert
1995). Kravtsov et al.
concluded that these density profiles are in reasonable agreement with 
those obtained in their high-resolution cosmological N-body simulations.
Other high-resolution simulations,
however, did not confirm the numerical results of KKBP98 
(Fukushige \& Makino 1997; Moore et al. 1998, 1999; Jing 1999), posing
a potential difficulty for the CDM models with respect to the
observations. The numerical, physical or observational analysis 
of the inner structure of galactic DM halos is beyond the scope of the 
present paper. Nevertheless, through our models, we shall explore 
the effects of a possible shallow 
core on the properties of the disc galaxies.

We shall introduce artificial shallow cores to our haloes. We look for
that the scaling parameters of the cores to be introduced are in rough
agreement with the rotation curves of low SB galaxies, taking into account
the contraction of their haloes {\it after} disc formation.
Since the density profile given by eq. (3) is in rough agreement with
that observed low SB rotation curves suggest, it is reasonable to use 
this profile for the galaxy halos. As a matter of fact, the shape of 
this profile is
similar to the typical density profile of our models (and to 
the NFW profile) except for the inner regions where 
$\rho (r) \propto r^{-0.2}$. Therefore, we deform {\it ad hoc}
the inner profile of our DM haloes by smoothly imposing the 
$\rho (r) \propto r^{-0.2}$ behavior from a given radius 
$r_{\rm core}=\nu r_{\max}$ down to the center, where $r_{\max}$
is the radius at $V_{\max}$ and $\nu <<1$. Because our approach is
evolutionary, we need to introduce the mentioned deformation at all
the times. Unfortunately, we only have information at $z=0$,
therefore, for all the other epochs we use $r_{\rm core}$ with the 
same $\nu $ defined at $z=0$.
The fittings applied by KKBP98 to low SB galaxies
and to the haloes obtained in their simulations show that the two 
scaling parameters of the profile (3), $r_0$ and $\rho _0$, are
linked. This allows us to roughly fix $r_0$ given other parameter 
as $\rho _0$ or $V_{\max}$ (see also Burkert 1995). With the formation 
of a disc ---even if this is of low surface density--- the 
original mass distribution of the dark halo 
changes. For instance, a model with the 
average MAH and $\lambda =0.08$ (low SB model) where the shallow core 
given by ec. (3) whith $r_0$ normalized to observations is introduced, {\it after} 
disc formation presents a too small $r_0$ for its $V_{\max}$. Now, 
if we increase $\nu $ for the initial halo, then at some value the
parameters $r_0$ and $V_{\max}$ of the system {\it after} disc formation will
agree with the parameters estimated for the observed low SB galaxies. 
The agreement occurs when $r_0$ increases roughly by a factor 1.4 with 
respect to the the KKBP98 estimates.


The main influence of the introduction of a shallow core in the DM
haloes appears on the dynamics of the
galaxy system. The same models presented in Figure 5 were again
calculated but with the inclusion of a shallow core in the DM
haloes. This core was introduced as described above: we deformed
our DM haloes in order their density profiles fitted eq.(3) with the
scaling parameters inferred by KKPB98 from rotation curves of low SB galaxies 
but making the core size ($r_0$) 1.4 times larger in order to ``return back''
the halo to its initial structure before disc formation. The obtained 
model rotation curve decompositions are plotted in
Figure 11. The disc component contribution to
the total rotation curve is now more significant than in the models
plotted in Fig. 5, and is in better agreement with several
observational and theoretical studies (see 
$\S 4.3$). The $V_d/V_t$ ratios for the catalog models of
$M_{\rm nom}=5\times10^{11}M_{\odot}$ calculated with the modified DM haloes
are plotted in Fig. 6 (empty circles). It is also important to note that 
the inclusion of a shallow core helps to obtain nearly flat rotation 
curves for models with small values of $\lambda $ and/or high values 
of $f_d$ that, otherwise, would have too steep rotation curves (see the end 
of $\S 4.2$). 

 
The inclusion of a shallow core in the DM haloes does not produce
considerable changes in the SSD, scale lengths, and gas fractions 
of the models. Some influence appears on $V_{\max}$. The 
$M_s-V_{\max}$ relation slightly shifts to the low velocity side (see
Figure 7, empty circles), being even in better agreement with the estimates 
from the observations than in the case of the models without core.
On average, the amplitude of this relation increases $\Delta $log $M_s
\approx 0.15$. The shallow core refers only to the most inner regions of 
the halo-galaxy system. Therefore, its influence on the  $V_{\max}$ is 
small because the  $V_{\max}$ of the original halo (without gravitational
contraction due to disk formation) is typically attained at radii larger 
than the disc size, i.e. far from the shallow core region.

Our models, which include the adiabatic contraction of
the DM due to disc formation, suggest that (i) the {\it original}
haloes of the present-day low SB galaxies had to have a shallow core
larger by roughly a factor 1.4 than that their rotation curves 
suggest according to KKBP98, and
(ii) the dynamics of normal (high SB) disc galaxies could be better
explained if the original DM haloes had such a core. The inner
structure of the galaxy dark haloes offers an important test for
structure formation theories (Moore 1994; Flores \& Primack 1994;
Burkert 1995; Moore et al. 1999). More observational efforts are
necessary in this direction.

\section{Summary and conclusions}

We have studied the formation and evolution of disc
galaxies in a $\Lambda $CDM$_{0.35}$ cosmology. We 
constructed a self-consistent model trying to avoid
free parameters. Our main
assumptions were (i) spherical symmetry and
adiabatic invariance during the gravitational collapse of the DM, (ii)
aggregation of the baryon matter to the disc in form of gas
(no merger) with a rate given by the cosmological aggregation
rate, (iii) detailed angular momentum conservation and adiabatic
invariance during the gas collapse, and (iv) 
stationarity and self-regulation of SF in the disc. The obtained
different density profiles of DM haloes and the dispersion agree with 
the results of cosmological N-body simulations (Avila-Reese et al. 1999).
The most typical profiles are well described by the NFW profile. The
properties of the model galaxies depend upon three initial factors: mass,
MAH, and spin parameter, $\lambda $. The results allow us
to predict and to understand several observational features of disc
galaxies:
 
{} (1) Within the observational and theoretical uncertainties, we find 
that the slope and zero-point of the TFR in the $I-$ and
$H-$bands may be directly determined by the cosmological initial conditions, 
principally the power spectrum of fluctuations of the CDM models.
The $M_s-V_{\max}$ relation (TFR) remains the same for different 
disc mass fractions $f_d$, in the cases where the disc makes a 
non-negligible gravitational contribution to the total rotation curve 
(when $f_d\grtsim 0.03$ for the cosmological parameters used here). 
The TFR can be used to normalize the power spectrum at galaxy
scales independently of uncertainties due to the assumed $f_d$.
$COBE$-normalized low density CDM models are favoured by this 
normalization.

{} (2) The rms scatter in the TFR, according to 
our models, is produced by the scatter in the halo formation histories
(MAHs) and by the dispersion of $\lambda$. As a result of compensating 
effects, the quadratic contribution of the latter is only a
25 per cent in the total rms scatter. Thus, a major contribution to the TFR 
scatter comes from the
stochastic nature of the protogalaxy MAHs. The total
scatter we obtain does not disagree with the observational
data, but, owing to the observational and theoretical uncertainties, it is
still premature to claim definitive conclusions.

{} (3) We explain why high and low SB galaxies show approximately the
same TFR, and why the slope of the correlation among
the residuals of the TF and luminosity-radius relations reported by
Courteau \& Rix (1998) is so small. We obtain similar results 
to those of Courteau \& Rix although the shape of the rotation 
curves of our models correlates 
with the SSD and the rotation curves are not
strongly dominated by the DM component. For models with a given total 
(star+gas) disc mass, as the SSD decreases, $V_{\max}$ decreases, {\it but}, 
owing to the dependence of the SF rate on the disc 
surface density, the stellar mass $M_s$ also decreases. Thus, models of
different surface density lie on the same relation in the $M_s-V_{\max}$
plane. Indeed, as our models and the observational data show, the disc
gas fraction ($f_g=M_g/(M_g+M_s)$) strongly correlates with the SSD (SB).

{} (4) For a given $f_d$, the shapes of the rotation curves are steeper 
as $\lambda $ decreases or as the MAH is more
active at early epochs (the DM halo is more concentrated). The
SSD depends upon $\lambda $ and the MAH, which explains why the shape
of the rotation curves depends upon the SSD, as the observations indicate.
If $f_d$ is too high ($\grtsim 0.08$), the rotation curves of
discs with small $\lambda $ ($<\lambda _{\min}\approx 0.04)$ decrease
too fast, and these discs are probably unstable. If $f_d$
is too small ($\lesssim 0.03$), then low and high SSD galaxies
will have very different TFRs, contrary to observations. 

{} (5) The rotation curve decompositions show a dominance of the 
DM component down to the very central regions for
most of the models with $f_d \approx 0.05$. This occurs because the inner
density profile of the DM haloes are steep ($\rho (r) \propto
r^{-1}$). The $V_d/V_t$ ratio increases with the SSD. On average, 
$V_d/V_t \sim 0.70$ for the high SSD models. 

{} (6) The discs in centrifugal equilibrium that form in the centre of
evolving CDM haloes with $\lambda $ constant in time, and with an
accretion rate dictated by the hierarchical MAH, have a nearly
exponential SSD distribution. The central SSD strongly depends on
$\lambda $, and less on the mass and the MAH.

{} (7) We have studied the effects a shallow core in the DM 
haloes would produce on the galaxy properties.
We found that the rotation curve decompositions of high SB 
galaxies agree better with the decompositions inferred from observations 
when the haloes have a shallow core. Because the 
rotation curve with a shallow core tends to be flatter, the minimum
possible value of $\lambda $ and the fraction of unstable discs 
decrease. On average, $V_d/V_t \sim 0.76$ for the high SSD models. 
The introduction of shallow cores slightly shifts
the models in the $M_s-V_{\max}$ plane towards lower velocities,
improving the agreement with the estimates inferred from the $I-$ and 
$H-$band TFRs.

The disc galaxy evolution models presented here make
use of several ingredients of the CDM-based hierarchical formation
scenario. We assumed that the hierarchical aggregation
of mass takes place as a gentle accretion process, discarding 
major mergers. Although a realistic galaxy formation model should take 
into account both accretion and merging, in the case of disc 
galaxies, the former could not have played an important role. One expects
that the aggregation of baryon matter was more uniform and gentle
than that of the DM due to its hydrodynamical properties and due to 
the re-heating at high redshifts (e.g. Blanchard, Valls-Gabaud,
\& Mamon 1992). Thus, even if the DM haloes suffered an active merging
process, luminous galaxies could have formed within large haloes in the 
way envisaged in the extended collapse picture. Our results showed
that several structural and dynamical properties of disc galaxies
and their correlations are closely related to the cosmological 
background, whereas other are consequence of evolutionary processes.
Among the former, the disc density and velocity profiles, and the
TFR are remarkable and they agree with the observations for 
the $\Lambda $CDM$_{\rm 0.35}$ universe
used here. However, we should also emphasize that the inner structure
of the CDM haloes and probably the scatter of the TFR, are in conflict
with observations. Both are probably associated with the statistical
nature of the primordial fluctuation field (Gaussian?) and/or to the
nature of the DM particles.

\section*{Acknowledgments}

We thank Julieta Fierro, Michael Richer, and Xavier Hern\'andez 
for helpful comments and for critically reading the original manuscript. 
We also thank S. Courteau for providing the data shown in Fig. 10a in 
electronic form and F. Angeles for computing assistance.
We are grateful to the announymous referee for comments
which were very helpful to improve the quality of the paper. 


\newpage

\begin{table}
  \caption{Average maximum velocities and
    scatters of the $M_h-V_{\max}$ relation for DM haloes.}
\label{tbl-1}
\begin{center}
 \begin{tabular}{|l|r|r|r|} \\ \hline
$M_v/M_{\odot}$ & 
$\overline{V}_m/kms^{-1}$ &
$\frac{\sigma _V}{\overline{V}}$ & $\sigma_ {M_h} (mag)$ \\ \hline
$4.2\times10^{10}$ & 55.3 & 0.101 & 0.33 \\ 
$3.8\times10^{11}$ & 112.0 & 0.093 & 0.31 \\ 
$3.6\times10^{12}$ & 221.0 &0.075 & 0.25 \\ \hline
 \end{tabular}
 \end{center}
 \end{table}

\begin{table}
  \caption{Average maximum velocities and
    scatters of the galaxy $M_s-V_{\max}$ relation.
\label{tbl-2}}
\begin{center}
 \begin{tabular}{|l|r|r|r|} \\ \hline
$M_s/M_{\odot}$ &
$\overline{V}_{m}/kms^{-1}$ &
$\frac{\sigma _V}{\overline{V}}$ &
$\sigma_ {M_s} (mag)$ \\ \hline
$9.6\times10^{8}$ & 68.4 & 0.110 & 0.38\\
$1.0\times10^{10}$ & 136.0 & 0.104 & 0.36\\
$1.1\times10^{11}$ & 270.0 & 0.095 & 0.33\\  \hline
 \end{tabular}
 \end{center}
 \end{table}

\clearpage
\begin{figure}
\resizebox{\hsize}{!}{\includegraphics{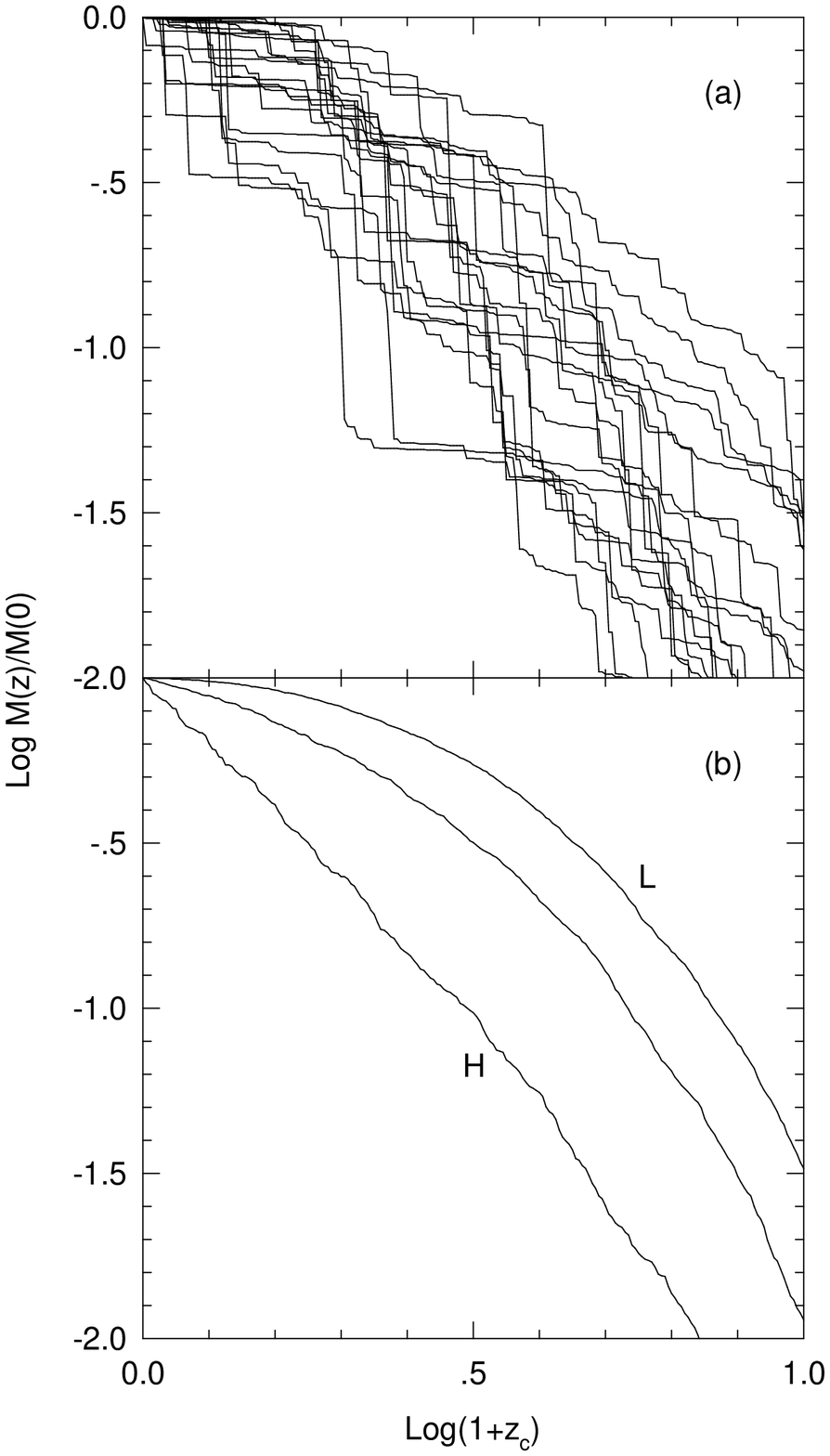}}
\caption[fig01.eps]{Mass aggregation histories
($M_{\rm nom}=5\times10^{11}M_{\odot }$) for a subsample of twenty
Monte Carlo realizations (a), and for statistically
selected cases over $10^4$ realizations (b). The central curve of panel (b) 
is the average MAH, while the two outer curves are averages over the 
extreme trajectories that
comprise $\sim 95$ per cent of all the trajectories (see text); L and H 
represent active early and very extended MAHs, respectively. The MAHs were
normalized to their respective $M_v$ at $z=0$; $M_v(0)$
slightly changes for each trajectory and has an average value of 
$M_v(0)=3.5\times10^{11}M_{\odot }$.}
\end{figure}

\begin{figure}
\resizebox{\hsize}{!}{\includegraphics{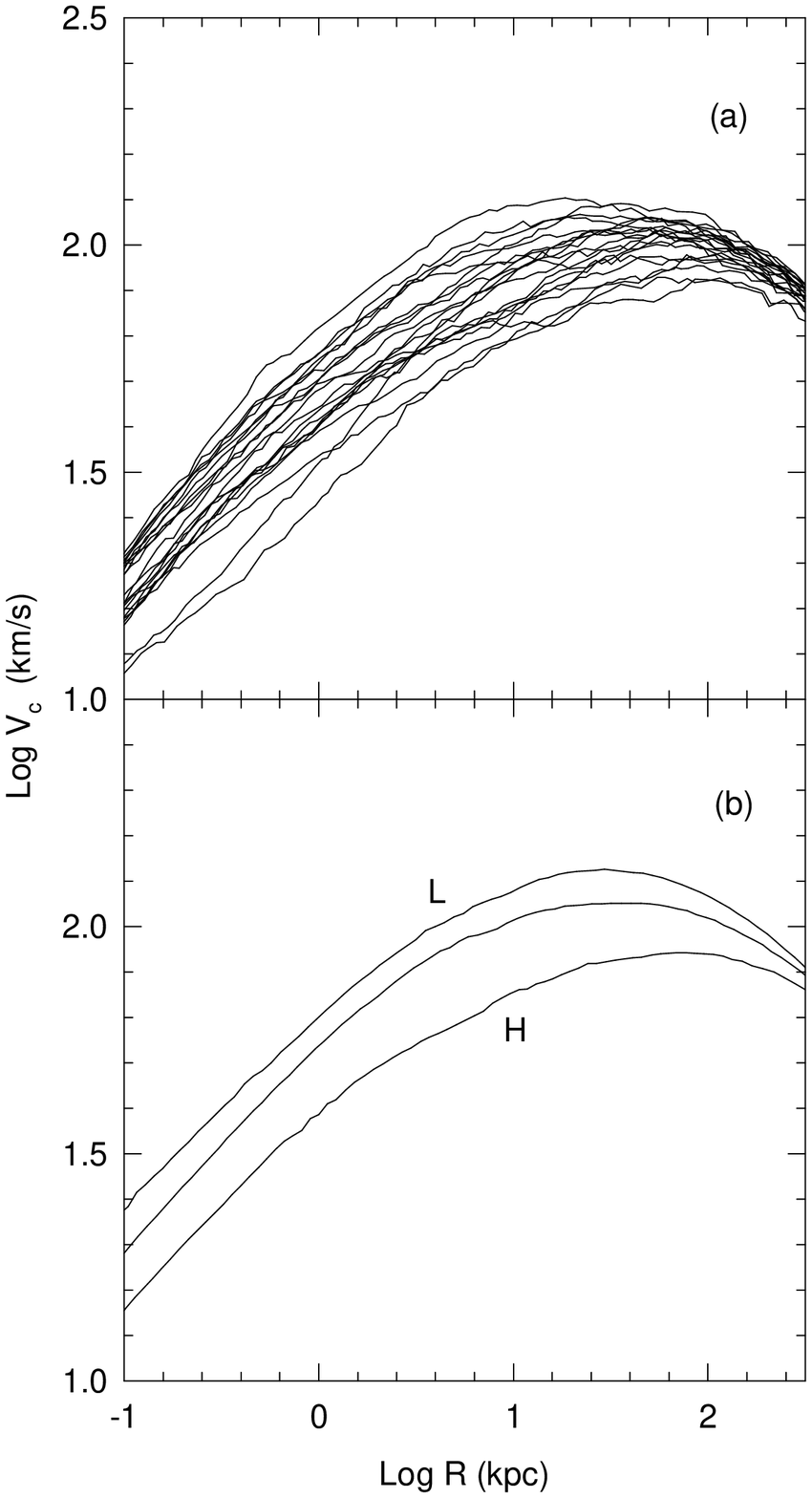}}
\caption[fig02.eps]{Circular velocity profiles for the DM haloes formed from
the same MAHs presented in panels (a) and (b) of Fig. 1, respectively.}
\end{figure}

\begin{figure}
\resizebox{\hsize}{!}{\includegraphics{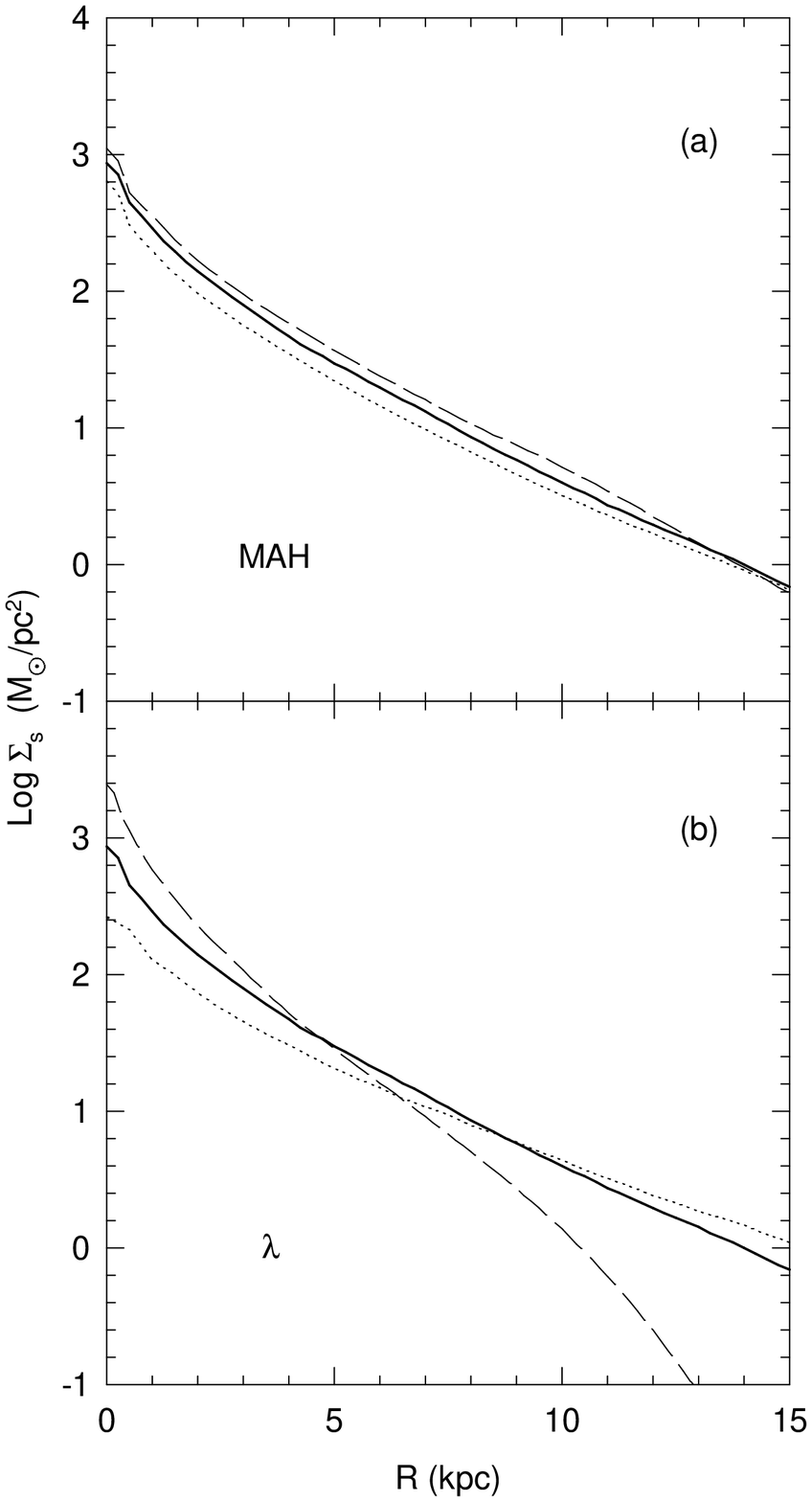}}
\caption[fig04.eps]{Stellar surface density profiles for models of 
$M_{\rm nom}=5\times10^{11}M_{\odot }$. In panel (a) the models correspond to
$\lambda =0.05$ and the average, L, and H MAHs (solid, dashed, and
dotted lines respectively). In panel (b) the models are for the
average MAH and  $\lambda =0.03, 0.05,$ and 0.08 (dashed, solid, and 
dotted lines respectively).}
\end{figure}

\begin{figure}
\resizebox{\hsize}{!}{\includegraphics{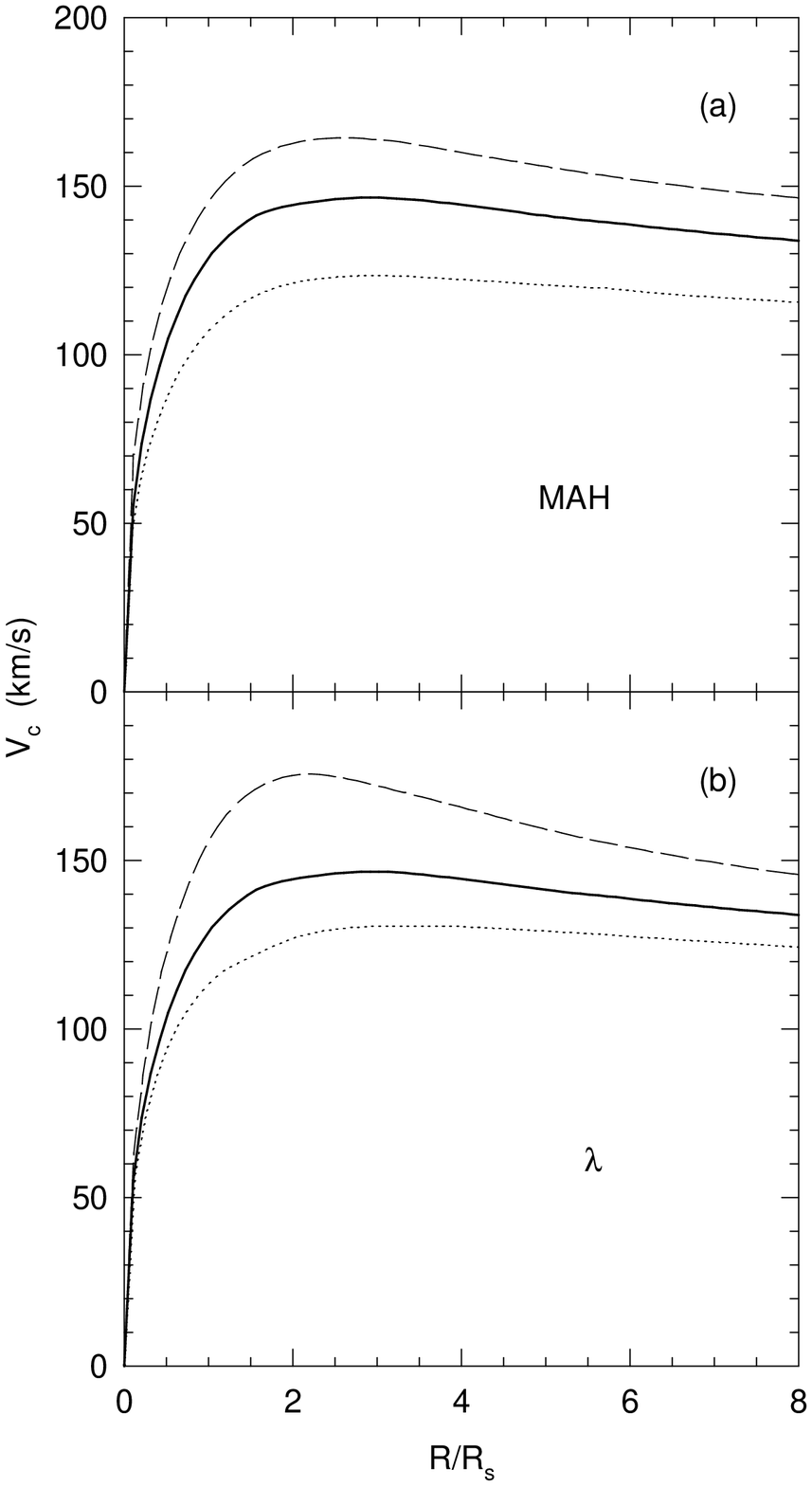}}
\caption[fig06.eps]{Total rotation curves for the same models
presented in Fig. 3. The radii are scaled to the corresponding
disc stellar scale lengths $R_s$.}
\end{figure}

\begin{figure}
\resizebox{\hsize}{!}{\includegraphics{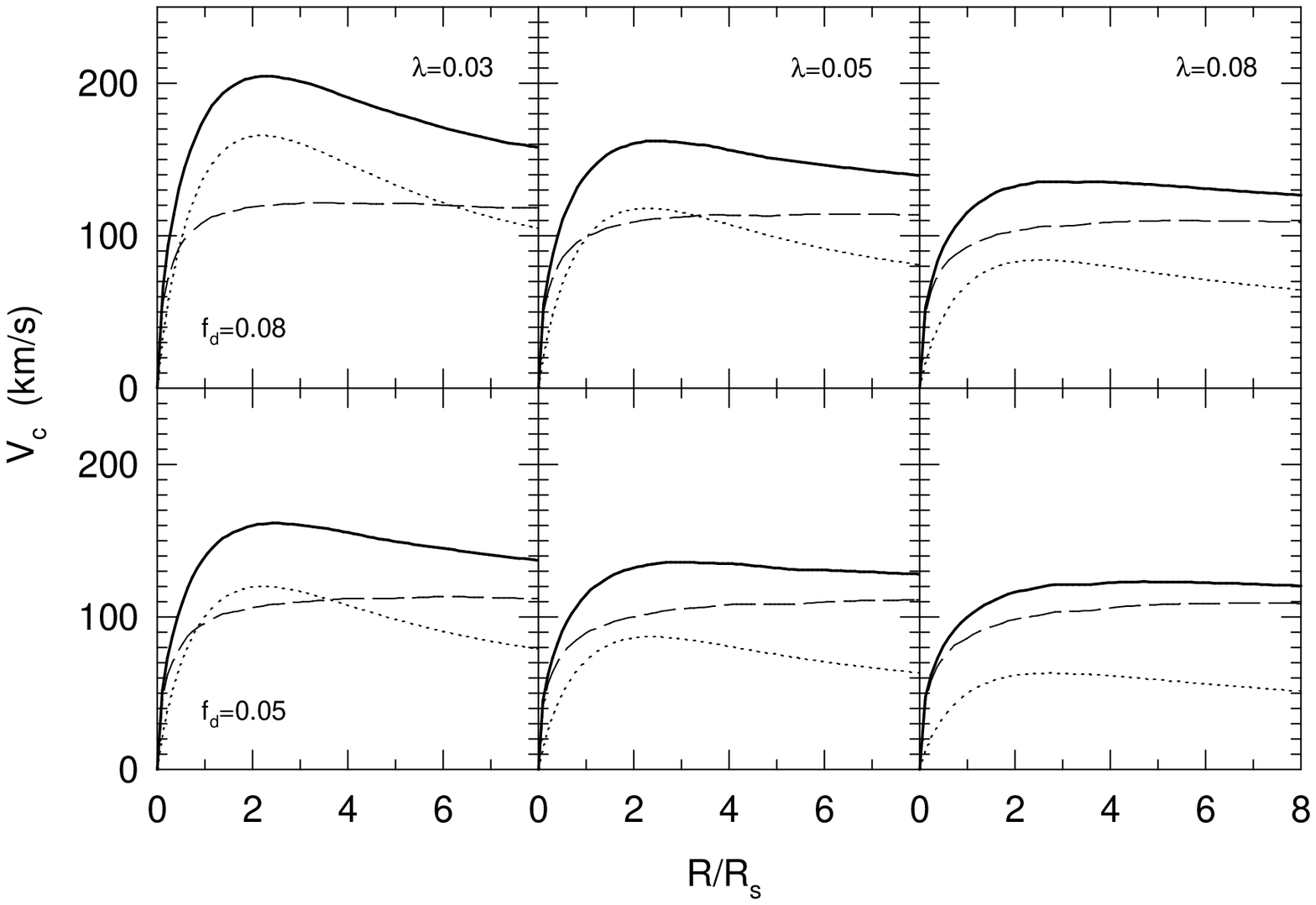}}
\caption[fig07.eps]{Rotation curve decomposition for several models of 
$M_{\rm nom}=5\times10^{11}M_{\odot }$ and with the average MAH. From left to
right the models are for $\lambda =0.03, 0.05,$ and 0.08,
respectively, and the upper and lower panels are for $f_d=0.08$ and
0.05. The dotted and dashed lines are the disc and DM halo
components, respectively, while the solid line is the total rotation
curve. The radii are scaled to the corresponding
disc stellar scale length $R_s$}
\end{figure}

\begin{figure}
\resizebox{\hsize}{!}{\includegraphics{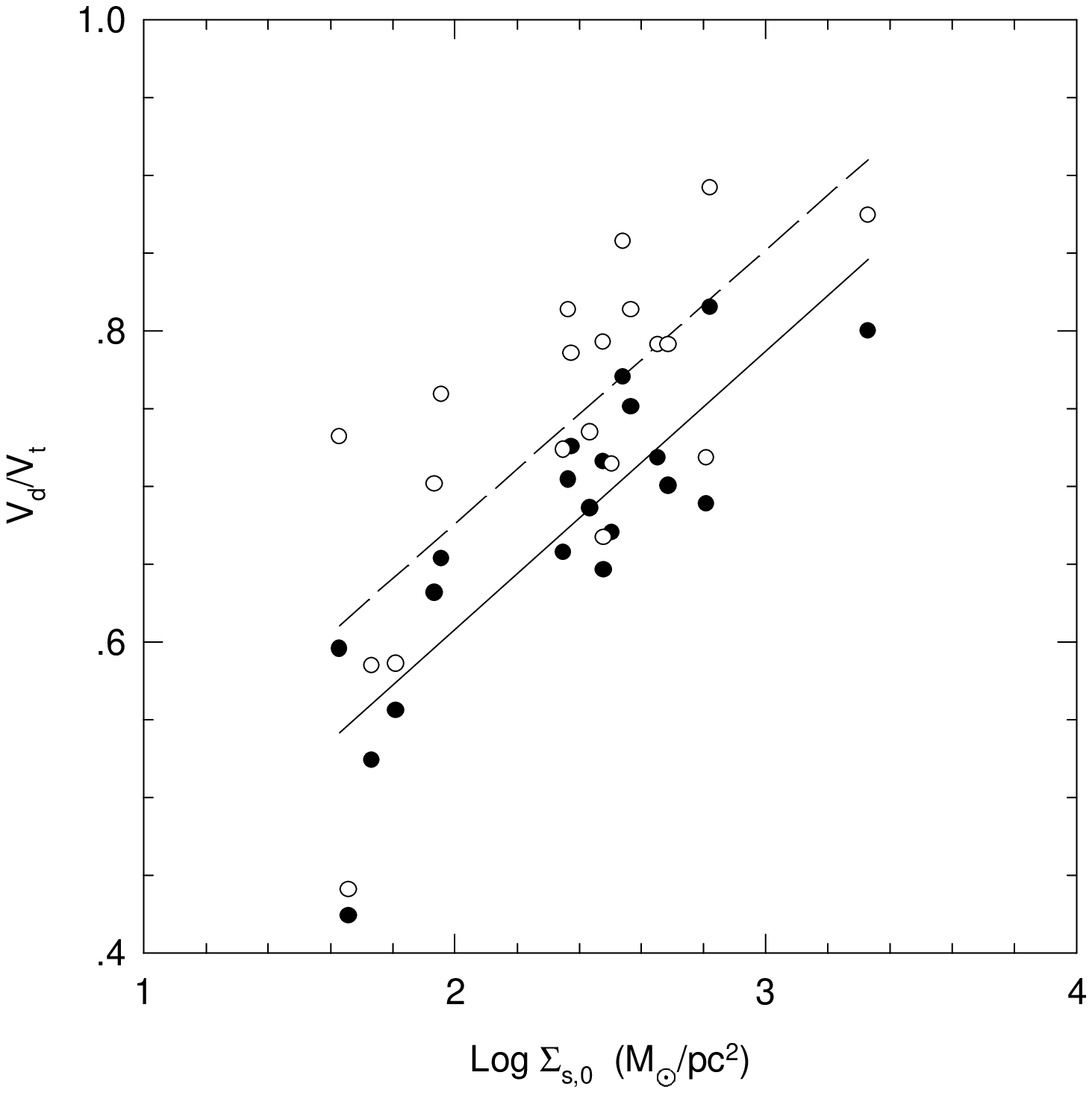}}
\caption[fig08.eps]{The maximum disc-to-maximum total velocity ratio
vs. the central SSD for a catalog ($M_{\rm nom}=5\times10^{11}M_{\odot }$) 
of twenty models (filled circles); the MAH and $\lambda $ of each
model is calculated from Monte Carlo simulations (see text). The
empty circles are for the same models but with a shallow core
artificially introduced in the DM haloes (see $\S 6$). The solid and
dashed lines are linear regressions for the filled and empty circles,
respectively.}
\end{figure}

\begin{figure}
\resizebox{\hsize}{!}{\includegraphics{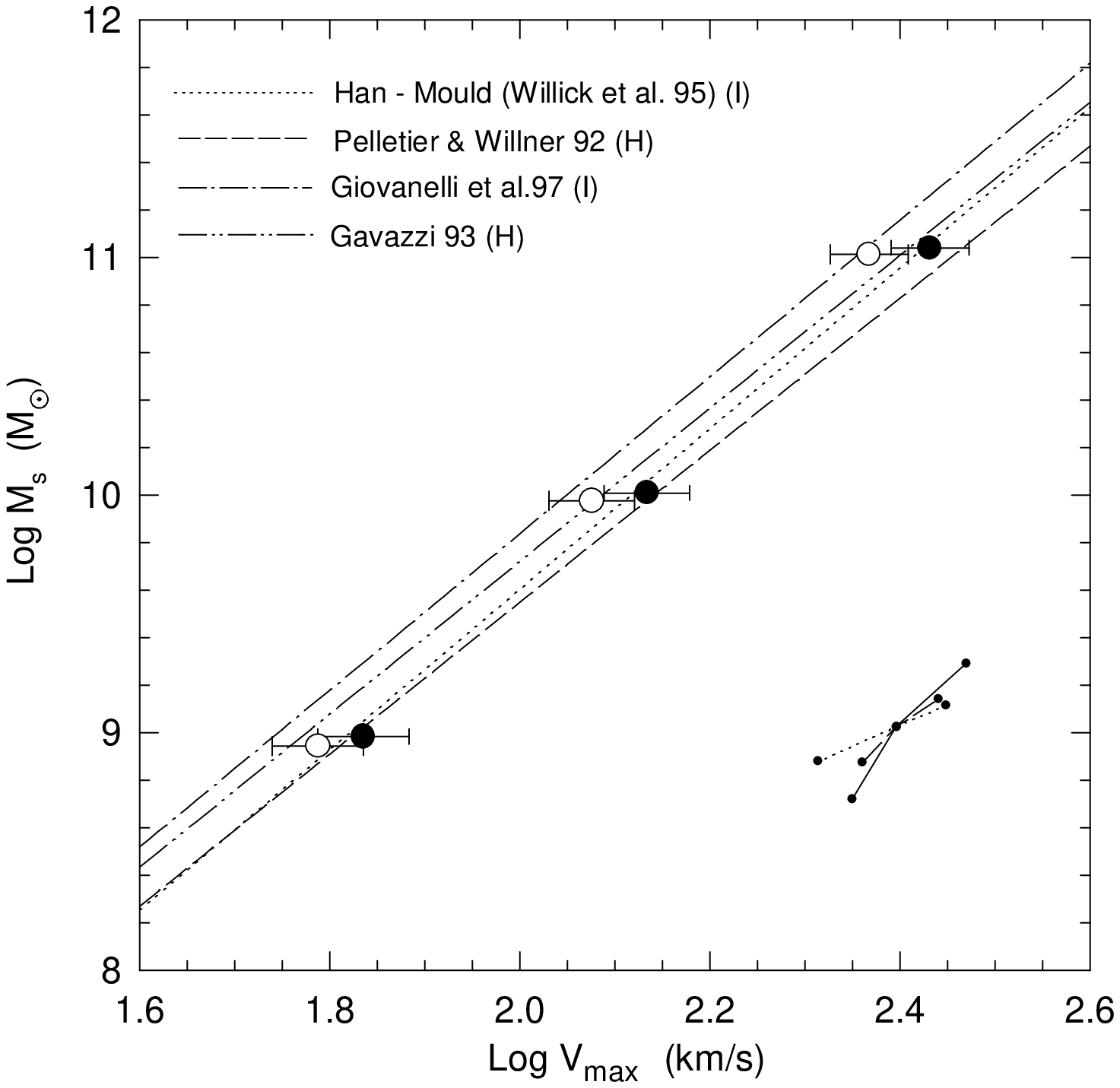}}
\caption[fig09.eps]{The $M_s-V_{\max}$ relation for models and
observations. The location of the models (filled circles) was
estimated from Monte Carlo catalogs calculated for different
masses. The empty circles are for the same catalog models,  
but with a shallow core artificially introduced in the DM haloes (see
$\S 6$). The error bars represent the standard deviations in the
velocity. The different lines correspond to the TFR given by several
observational studies (indicated in the panel). To convert the $I-$ and
$H-$band luminosities to stellar masses we have used 
$\Upsilon _I=1.8\left( \frac{M_s}{5\times 10^{10}M_{\odot }}\right) ^{0.07}h$,
and $\Upsilon _H=0.55$, respectively ($h=0.65$ was assumed in the latter case).
The small dots and lines in the right lower corner represent
several particular models of  $M_{\rm nom}=5\times10^{11}M_{\odot
}$. From left to right the continuous line connect models of 
$f_d=0.03, 0.05,$ and 0.08 with the average MAH and $\lambda =0.05$; the 
dotted line connect models of L, average and H MAHs with $f_d=0.05$ and
$\lambda =0.05$; and the dashed line connect models of $\lambda =0.08,
0.05,$ and 0.03 with the average MAH and $f_d=0.05$. All these models where
shifted down to the right in order to avoid overlapping. The shifts of the models
when changing $f_d$, the MAH, and $\lambda$  are practically
independent of the mass.} 
\end{figure}
 
\begin{figure}
\resizebox{\hsize}{!}{\includegraphics{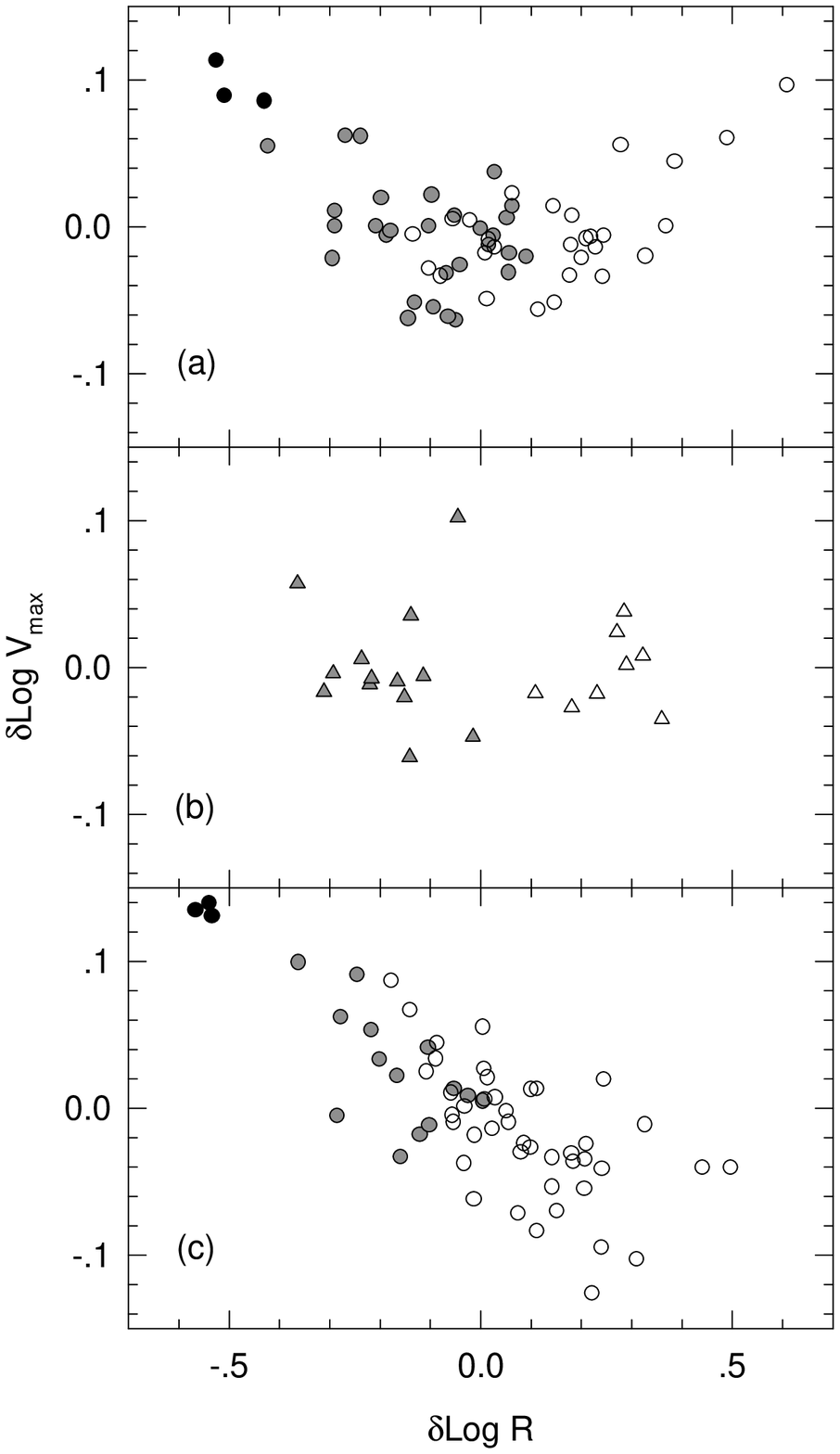}}
\caption[fig10.eps]{Correlations among the residuals of (a) the
$M_s-V_{\max}$ and $M_s-R_s$ relations from catalog models of different 
masses, (b) the K'-band TFR and the luminosity-scale length relation
from observational data of the Ursa Mayor cluster galaxies (Verheijen
1997), and (c) the $M_{\rm tot}-V_{\max}$ and $M_{\rm tot}-r_{\rm
tot}$ relations from
the same models of panel (a); $M_{\rm tot}$ and $r_{\rm tot}$ are the disc
total (gas+stars) mass and scale length, respectively. The model data
(panels (a) and (c)) were divided into 3 groups: high, normal, and low
surface density galaxies (black, gray, and empty circles,
respectively), according to their stellar (a) or total (c) surface
densities ($\Sigma _0>2000, 2000<\Sigma _0<200$, and $\Sigma _0<200$,
respectively, where $\Sigma _0$ is the central stellar or total
surface density). The observational data (panel b) were divided into high and
low surface brightness galaxies (gray and empty triangles,
respectively); the threshold in the central SB used for the division 
was $\mu _0=18.5$ K'-mag/arcsec$^{-2}$ (see Verheijen 1997).}
\end{figure}
 
\begin{figure}
\resizebox{\hsize}{!}{\includegraphics{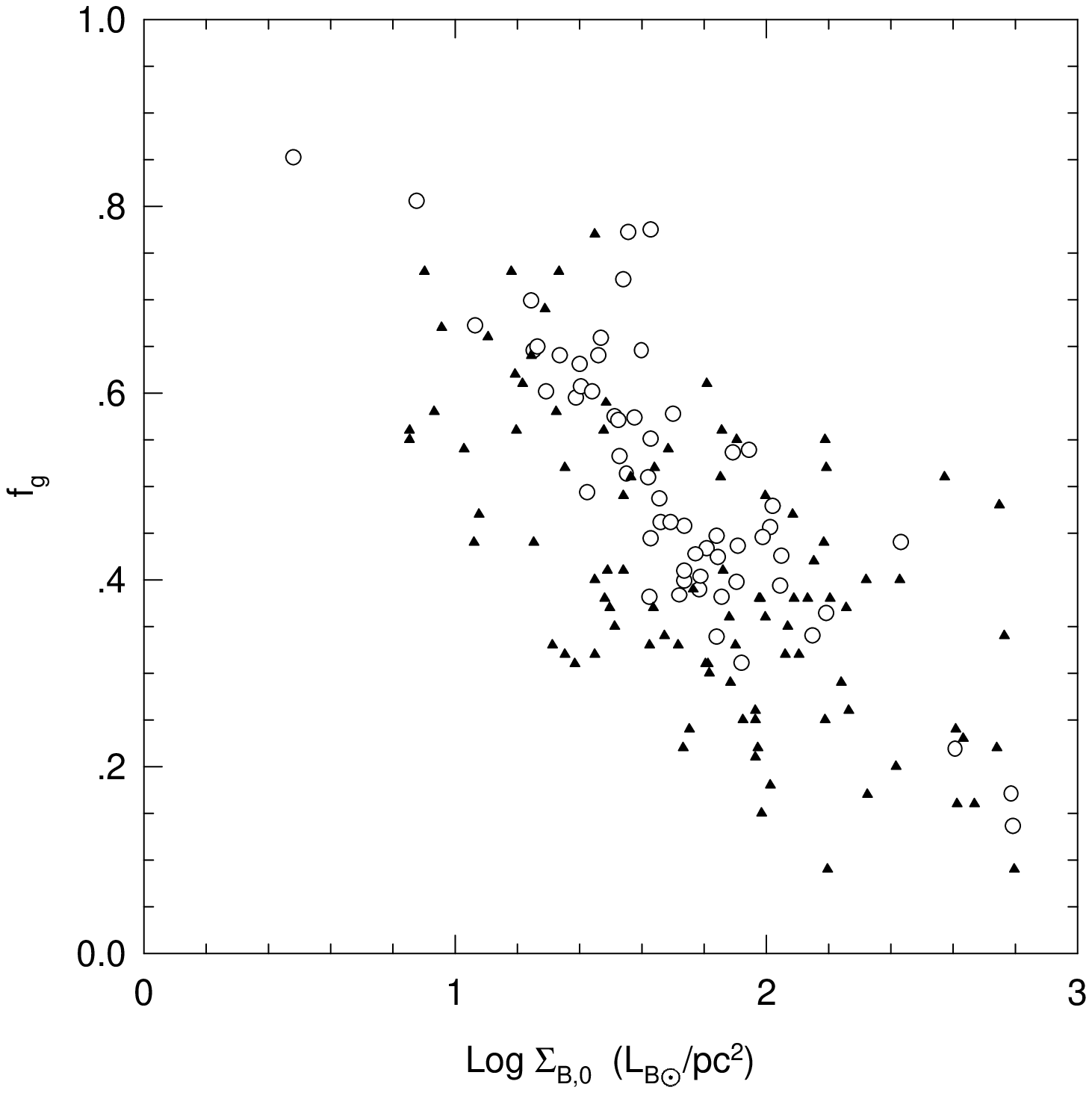}}
\caption[fig11.eps]{The final gas mass fraction
($f_g=M_{gas}/(M_{gas}+M_s))$) vs. the central B-band SB of models
(circles) and observations (triangles). The models are based upon the same
Monte Carlo catalogs used in Fig. 7 (without cores). The observational 
data were taken from de Blok \& McGaugh 1997.}
\end{figure}
 
\begin{figure}
\resizebox{\hsize}{!}{\includegraphics{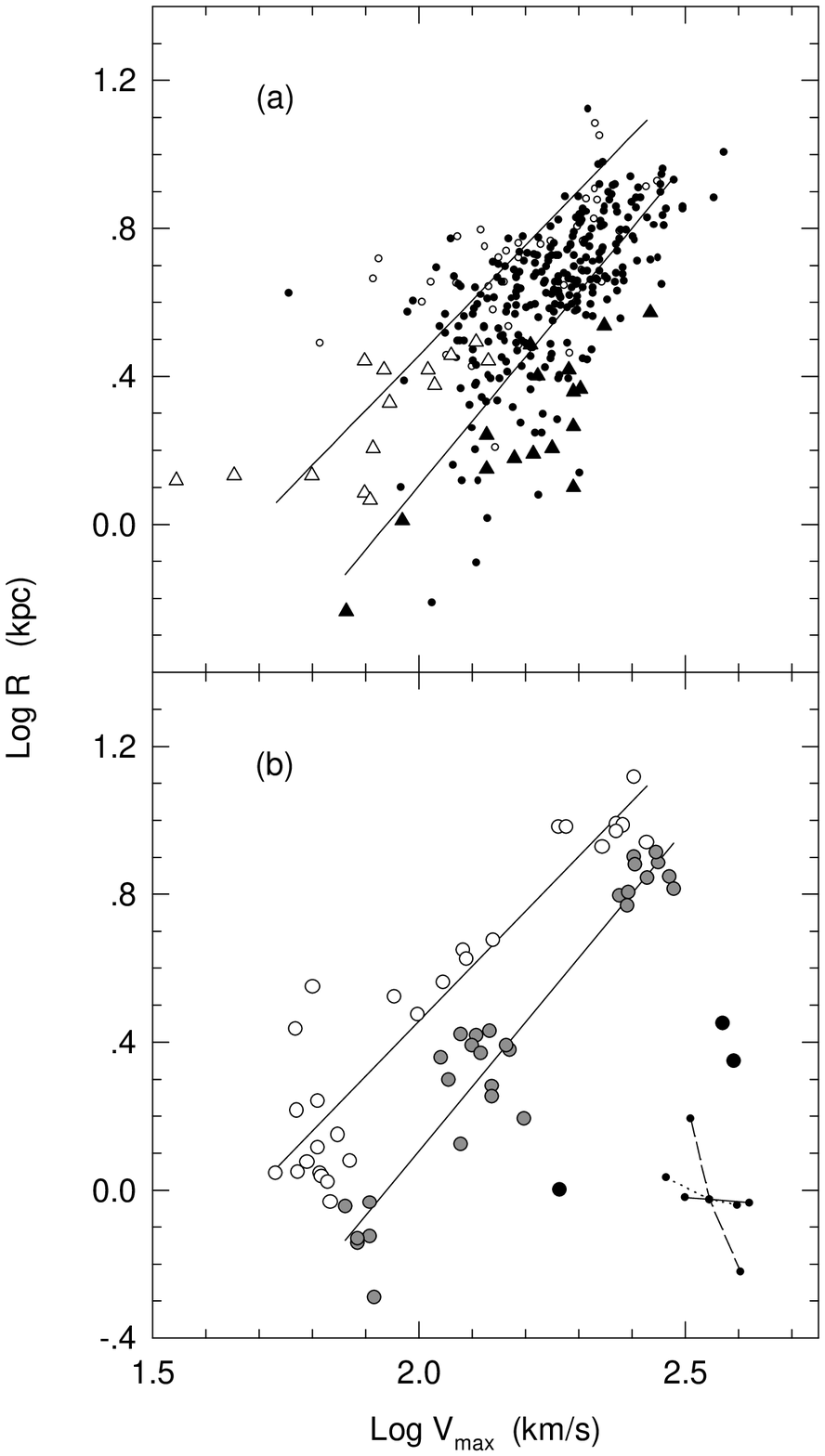}}
\caption[fig12.eps]{Scale length vs. $V_{\max}$ for the same models
described in Fig. 7a (b), and for observational data (a). The dots in
panel (a) are from Courteau (1996, 1997), and the triangles are from
Verheijen (1997). Both observational samples were divided into low and 
high SB galaxies (empty and filled symbols, respectively), according
to the criteria given in the text. The solid lines are the linear
regressions to the low (upper) and normal (lower) SSD model galaxies of 
panel b. The dots in the right lower corner of panel (b) represent
several particular models of $M_{\rm nom}=5\times10^{11}M_{\odot
}$(see text ); these models where shifted by $-0.43$ in log 
$R_s$ and by 0.41 in log $V_{\max}$ in order to avoid overlapping. The 
shifts are roughly the same for other masses.}
\end{figure}
 
\begin{figure}
\resizebox{\hsize}{!}{\includegraphics{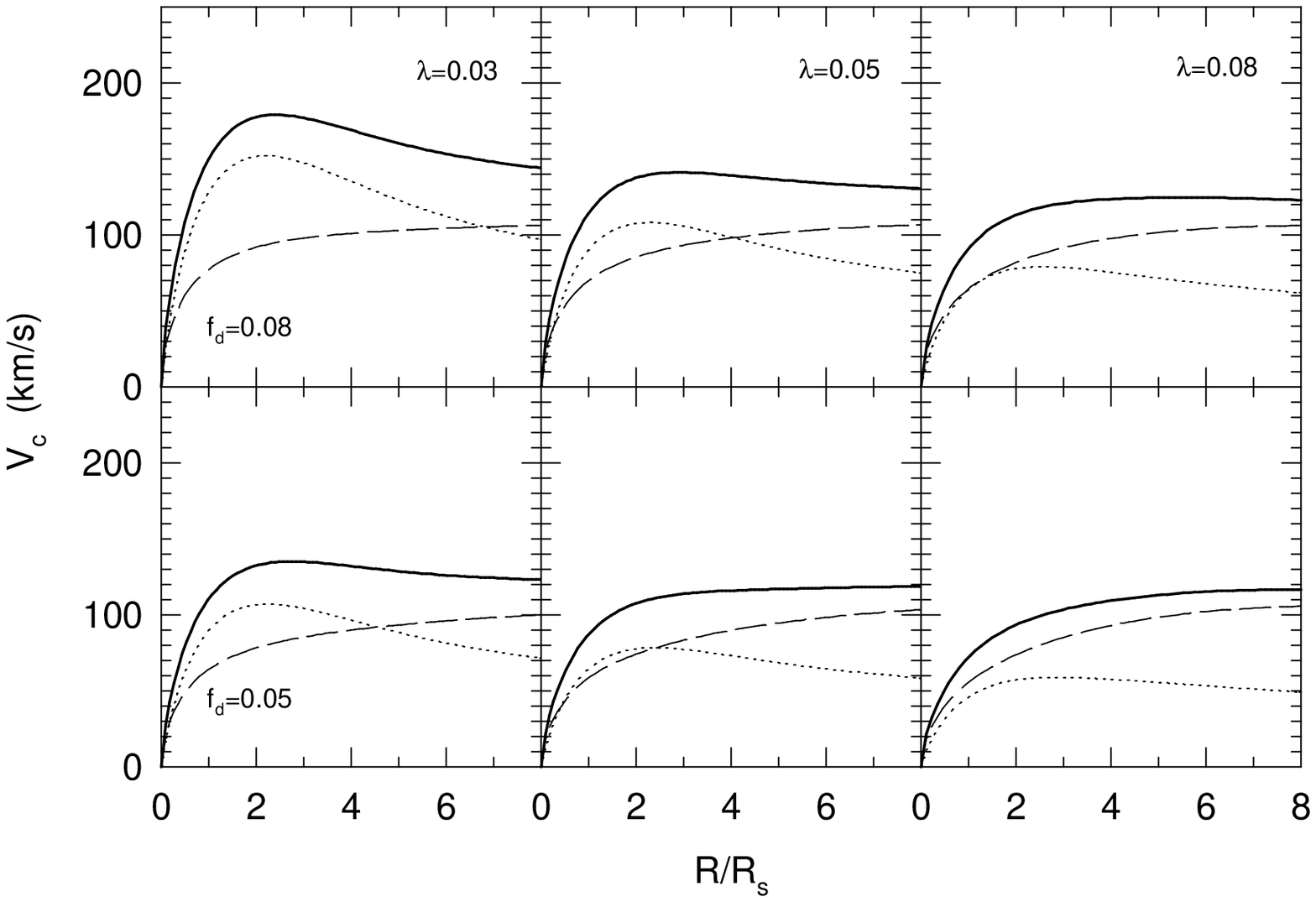}}
\caption[fig14.eps]{Same as Figure 5 but with a shallow core
artificially introduced in the DM haloes (see text).}
\end{figure}

\end{document}